\documentclass[reprint,prb,superscriptaddress,amsmath,amssymb]{revtex4-2}
\usepackage{graphicx}
\usepackage{hyperref}
\usepackage{dcolumn}% Align table columns on decimal point
\usepackage{color}
\usepackage{gensymb}
\usepackage{hyperref}
\usepackage[dvipsnames]{xcolor}

\usepackage{nicefrac}
\definecolor{nicegreen}{RGB}{25,120,30}

\newcommand{\lanio}{La$_2$NiO$_4$}
\newcommand{\ketxyu}{\ensuremath{\left|xy,\uparrow\right\rangle}}
\newcommand{\ketyzu}{\ensuremath{\left|yz,\uparrow\right\rangle}}
\newcommand{\ketzru}{\ensuremath{\left|3z^2\!-\!r^2,\uparrow\right\rangle}}
\newcommand{\ketxzu}{\ensuremath{\left|xz,\uparrow\right\rangle}}
\newcommand{\ketxxu}{\ensuremath{\left|x^2\!-\!y^2,\uparrow\right\rangle}}
\newcommand{\ketxyd}{\ensuremath{\left|xy,\downarrow\right\rangle}}
\newcommand{\ketyzd}{\ensuremath{\left|yz,\downarrow\right\rangle}}
\newcommand{\ketzrd}{\ensuremath{\left|3z^2\!-\!r^2,\downarrow\right\rangle}}
\newcommand{\ketxzd}{\ensuremath{\left|xz,\downarrow\right\rangle}}
\newcommand{\ketxxd}{\ensuremath{\left|x^2\!-\!y^2,\downarrow\right\rangle}}
\newcommand{\braxyu}{\ensuremath{\left\langle{}xy,\uparrow\right|}}
\newcommand{\brayzu}{\ensuremath{\left\langle{}yz,\uparrow\right|}}
\newcommand{\brazru}{\ensuremath{\left\langle{}3z^2\!-\!r^2,\uparrow\right|}}
\newcommand{\braxzu}{\ensuremath{\left\langle{}xz,\uparrow\right|}}
\newcommand{\braxxu}{\ensuremath{\left\langle{}x^2\!-\!y^2,\uparrow\right|}}
\newcommand{\braxyd}{\ensuremath{\left\langle{}xy,\downarrow\right|}}
\newcommand{\brayzd}{\ensuremath{\left\langle{}yz,\downarrow\right|}}
\newcommand{\brazrd}{\ensuremath{\left\langle{}3z^2\!-\!r^2,\downarrow\right|}}
\newcommand{\braxzd}{\ensuremath{\left\langle{}xz,\downarrow\right|}}
\newcommand{\braxxd}{\ensuremath{\left\langle{}x^2\!-\!y^2,\downarrow\right|}}
\newcommand{\st}{\mathrm}
\usepackage[normalem]{ulem}

\begin{document}

\author{R.O. Kuzian}
\affiliation{Institute for Problems of Materials Science NASU, Krzhizhanovskogo 3, 03180 Kiev, Ukraine}
\affiliation{Aix-Marseille Universit\'e, IM2NP-CNRS UMR 7334, 
Campus St. J\'er\^ome, Case 142, 13397 Marseille, France}

\author{O. Janson}
\affiliation{Leibniz Institute for Solid State and Materials Research IFW Dresden, 01171 Dresden, Germany}

\author{A. Savoyant}
\affiliation{Aix-Marseille Universit\'e, IM2NP-CNRS UMR 7334, 
Campus St. J\'er\^ome, Case 142, 13397 Marseille, France}

\author{Jeroen van den Brink}
\affiliation{Leibniz Institute for Solid State and Materials Research IFW Dresden, 01171 Dresden, Germany}

\author{R. Hayn}
\affiliation{Aix-Marseille Universit\'e, IM2NP-CNRS UMR 7334, 
Campus St. J\'er\^ome, Case 142, 13397 Marseille, France}
\affiliation{Leibniz Institute for Solid State and Materials Research IFW Dresden, 01171 Dresden, Germany}
\affiliation{Max-Planck Institute for Physics of Complex Systems, N\"othnitzer Str., 01187 Dresden, Germany}

%\title{\emph{Ab initio} ligand field theory}

\title{ {\emph{Ab initio} ligand field approach to determine electronic multiplet properties} }

\begin{abstract}
A method is developed to calculate the ligand field (LF) parameters and the multiplet spectra of local magnetic centers with open
$d$- and $f$-shells in solids in a parameter-free way. 
This method proceeds from density functional theory and employs Wannier projections of nonmagnetic band structures 
onto local $d$- or $f$-orbitals.
Energies of multiplets and optical, as well as X-ray spectra are determined by exact numerical diagonalization of a local
Hamiltonian describing Coulomb, LF, and spin-orbit interactions. The method is tested for several 3$d$- and 5$f$-compounds
for which the LF parameters and multiplet spectra are experimentally well known. In this way, we obtain good 
agreement with experiment for La$_2$NiO$_4$,
CaCuO$_2$, Li$_2$CuO$_2$, ZnO:Co, and UO$_2$.

%{\color{red} (RH: to be completed.)} We propose a relatively simple DFT-based approach to the calculation of 
%$d$-level splitting in transition metal compounds.
%The approach is applied to various compounds.
\end{abstract}

\date{\today}

\maketitle
\section{introduction} 

A fundamental problem in electronic structure theory of solids is the proper description of multiplet effects
of local magnetic centers built up of $d$ or $f$ electrons, which are intrinsically many-body states, in translational invariant settings.
Electronic structure calculations 
based on density functional theory (DFT) are successful in predicting the atomic positions, the electronic, magnetic, and state densities in an
\emph{ab-initio} manner with high precision. However, these calculations use a mean-field potential and one-electron 
states, and  thus in principle cannot access the many-electron multiplet levels characterized 
by strong Coulomb interactions, electron correlations, and spin orbit coupling. 
However, the structure of local multiplets is well understood since many years in atomic physics, mainly based on 
group theory applied to open-shell atoms or ions. Such multiplets can persist in solids, either as sharp 
levels in the 
gap of insulators or semiconductors or as resonances in metals and small gap semiconductors. 
The difference to the case of isolated atoms and ions is the appearance of a small number of new parameters 
which effectively describe 
the influence of the surrounding crystal.
 Traditionally, they are often called crystal field (CF)
\cite{Bethe1997} parameters since they are partially
caused by the electrostatic Madelung potential in the crystal. However, in most cases, and 
especially in the examples 
of 3$d$ and 5$f$ compounds we are going to treat, the hybridization to the 
neighboring ligands, also known 
as ligand fields (LF), makes a much bigger influence and that is the term which we will prefer 
here. Without explicit hybridization with neighboring orbitals, a given set of LF parameters 
determines entirely the influence of the environment on the multiplet structure of 
a localized open $d$- or $f$-shell. 

The knowledge of LF parameters and multiplet spectra is especially important for strongly correlated 
systems being of high actual interest 
since they may show 
superconductivity
\cite{Bednorz1986}, multiferroism  \cite{Spaldin2019,Reschke2020}, or spin liquid behaviour \cite{Broholm2020,li20ncomms,liu20prl}, 
as well as many other interesting properties. 
Recently, new experimental techniques like resonant inelastic X-ray scattering 
(RIXS) lead to important improvements to measure multiplet spectra for cuprates,  
nickelates \cite{Sala2011,Fabbris2017} and other materials. 

%One short paragraph about 3d and 5f materials and experimental investigations ... 
%strong correlation physics supraconductivity ... orbital ordering .... exotic states of matter ....
%.... multiferroics ...
% applications ....  
% ... models are constructed ...
% DMS .... nuclear fuel.... LF parameters are important to characterize the local magnetic centers 
%.... experiments EPR ... neutron scattering .... RIXS .... and many other }

In the literature, one can find several approaches to calculate LF or CF parameters. First of all, there are wave 
function quantum chemistry methods \cite{Hozoi2011}.  
However, it is difficult for these methods to treat a periodic crystal and they 
%It is, however, difficult to include properly a small cluster into the periodic crystal  
%and wave function methods 
become numerically expensive for heavy ions and large systems. 
%DFT methods, on the 
%other hand, have the problem to account for self-interaction. 
This motivates attempts
 to calculate multiplets and LF parameters in an ab-initio
style and based on DFT \cite{Steinbeck1996,Novak13}. 
Since we are dealing here with highly correlated systems it is 
tempting to start with the local spin density (LSDA) or spin dependent generalized gradient 
approximation (SGGA) corrected for Hubbard
correlation effects, i.e. the LSDA+$U$ or SGGA+$U$ functionals. 
And indeed the CF parameters of lanthanide and actinide
dioxides were successfully calculated by Zhu and Ozoli\c{n}\v{s} \cite{Zhou2012} based 
on the LSDA+$U$ method with occupation matrix control. 
However, it is necessary in that case to correct the LSDA+$U$ functional for self-interaction and 
double counting terms and 
to modify it. Instead, we here use a simpler method by starting with the original non spin-polarized 
GGA functional \cite{PBE96}.
Similar to the approach of Ref.\  \onlinecite{Reschke2020}, 
we obtain then the LF parameters by a Wannier fit to the non-magnetic band structure. 
%{\color{blue} \sout{But we will
%not restrict ourselves to determine solely the LF parameters}}. {\color{blue} 
Further, in a second step, we use the LF parameters in an exact
diagonalization computer program 
to predict the outcome of a multitude of experiments being
sensible to local multiplet effects, i.e. electron paramagnetic resonance (EPR), optical 
spectroscopy, inelastic neutron scattering (INS), X-ray absorption and 
X-ray magnetic circular dichroism (XAS and XMCD) as well as resonant inelastic X-ray scattering (RIXS).
That computer program ELISA (Electrons Localized In Single Atoms) was successfully used before to analyze the XAS 
and XMCD spectra of Mn-based metal-organic networks \cite{Giovanelli14}  so that we will concentrate here on 
the other experimental techniques. 
In the present publication,
we test our method for La$_2$NiO$_4$, CaCuO$_2$, Li$_2$CuO$_2$, Co impurities in ZnO, and UO$_2$.

In our approach, we calculate the influence of the neighboring ligands on the local one-electron energy levels of 
the $d$- or $f$-center, i.e. the LF parameters. We then diagonalize exactly the local, atomic like, 
multiplet Hamiltonian  and do not treat the ligand $p$-orbitals explicitly. That is different to the ab-initio multiplet 
LF theory of Haverkort et al. \cite{Haverkort2012}. Our approach is 
simpler, and yet fully justified as long as
charge-transfer processes in optics or X-ray spectroscopy are not at play. 
Finally, we obtain various \emph{ab initio} simulations of multiplet spectra without any 
adjustable parameters others than the line width. 
Calculating the LF parameters has also the 
advantage to establish a connection to traditional crystal field methods with parameters obtained by 
fitting to experimental data.

\section{Method}

The multiplet spectrum of a local $d$- or $f$-center depends on the local Coulomb interaction, the spin-orbit coupling 
and the LF parameters. The Coulomb and spin-orbit parameters of the free ion can either be obtained by Hartree-Fock
calculations \cite{Groot2005,Groot2010} or by fitting the optical spectra of free ions. These data 
are easily accessible in the NIST data set \cite{NIST_ASD} and that is the way
we follow here. The Coulomb parameter are usually slightly screened in the solid but we will show that this screening 
has only little influence on the multiplet spectrum, especially in the low-energy part we are mostly interested in.

In a second step, we perform GGA calculations to obtain the non spin-polarized band structure.
%First step: Coulomb and spin-orbit parameters from spectra of the free ion (NIST data)
%Second step: GGA calculations (non-magnetic) and Wannier fit
%(scalar relativistic: only LF parameter, full relativistic: allows also the determination of the spin-orbit coupling)
%Third step: Multiplet calculations using the ELISA code
%Applications: Calculation of different kinds of spectra: EPR, optical spectra, XAS and XMCD, RIXS.
These density functional theory calculations were performed here using the
full-potential local-orbital (FPLO) code \cite{FPLO,Koepernik99}. 
We have used the default FPLO basis. 
The exchange and correlation potential of Ref.\ \cite{PBE96} was employed, i.e. the generalized gradient 
approximation (GGA) functional. 
Wannier functions (WF) were constructed via projection onto the  
respective $d$- or $f$-orbitals of a magnetic atom; the 
site symmetry is fully taken into account; 
as it is implemented in the FPLO code \cite{Eschrig2009}.
We first do this "Wannier fit" for the scalar relativistic case and obtain then the LF parameters by a fit to the on-site energies
of the Wannier expansion. 
%{\color{blue} ( Roman or Oleg, maybe it will be good to insert here few words about the wannier fit limitation: small overlapp of 3d orbitals with other orbitals, locaization, etc... What do you think? RH: Usually, the limitations of the method are discussed in the end of the article.)}
Next, we perform the same procedure for the full relativistic, but nonmagnetic GGA
functional. That does not change the LF parameters, but opens the possibility to calculate also the spin-orbit parameter
being reduced in the solid with respect to the free ion value. A remarkable reduction was observed
for the cuprates which we investigate and will be discussed in detail
later on.

The final step of our method consists in the exact diagonalization of an electronic 
Hamiltonian within the finite state space of open atomic shell(s)
using the ELISA code. 
That code is able to treat several electronic configurations, an arbitrary number 
of electrons in each shell, and contains all the Coulomb interactions within one shell and  between different shells. 
The influence of the surrounding ligands is taken into account by a  LF Hamiltonian written in terms of 
Steven's operators with appropriate symmetry. The Hamiltonian contains Coulomb interaction, the ligand field, the spin-orbit coupling, and the magnetic field: 

\begin{equation}\label{H}
\hat{H}=\hat{H}_\st{Coul} + \hat{H}_\st{LF} + \hat{H}_\st{SO}+ \hat{H}_\st{B} \; .
\end{equation}
In the examples treated in the present study we are going to consider $2p$, $3d$ and 5$f$ shells. 
To describe X-ray transitions we have to consider at least two different configurations. The Coulomb interaction 
\begin{equation}
\hat{H}_\st{Coul}=\frac{1}{2} \sum_{m_i   \sigma  \sigma' } 
  V_{m_1 m_2 m_3 m_4}   c^{\dagger}_{m_1 \sigma} 
 c^{\dagger}_{m_2 \sigma'} c_{m_3 \sigma'} c_{m_4 \sigma}
\end{equation}
is treated as in rotationally invariant atoms and parametrized by Slater parameters.
For instance, the Coulomb interaction in the $3d$ shell is given by
%\begin{eqnarray}
%V_{m_1 m_2 m_3 m_4} = 25 \sum_{k=0,2,4} (-1)^{m_1+m_4} F^{(k)} \cdot \nonumber \\
%\cdot C^{k,0}_{2,2,0,0} C^{k,0}_{2,2,0,0} 
%C^{k,m_3-m_1}_{2,2,m_3,-m_1} C^{k,m_4-m_2}_{2,2,m_4,-m_2} 
%\end{eqnarray}
%
%{\color{blue}
%
%(I would change the Eq. 3 as follow:)
\begin{eqnarray}
V_{m_1 m_2 m_3 m_4} = 25 \sum_{k=0,2,4} (-1)^{m_1+m_4} F^{(k)}  \nonumber \\
\times \left[ C^{k,0}_{2,2,0,0} \right]^2 
C^{k,m_3-m_1}_{2,2,m_3,-m_1} C^{k,m_4-m_2}_{2,2,m_4,-m_2} 
\end{eqnarray}
with three Slater parameters $F^{(k)}$ which can also be expressed by three 
Racah parameters in the form 
$F^{(0)}=A+\frac{7}{5}C$, 
$F^{(2)}=49 B + 7 C$ and $F^{(4)}=\frac{441}{35}C$. $C^{k,m}_{l_1,l_2,m_1,m-m_1}$ are the usual 
Clebsch-Gordon parameters in the
$3d$ shell when $l_1=l_2=2$. A similar expression holds also for the Coulomb interaction in  
other shells and 
inbetween 
different shells.
The crystal environment influences the spectra by the electrostatic Madelung potential (crystal field) but 
also by hybridization to the 
neighboring ligands. Since the second contribution is probably dominant in most cases we prefer the term ligand field
Hamiltonian here. It is a non-interacting Hamiltonian, like also the spin-orbit coupling and the Zeemann term, which can be written as
\begin{equation}\label{H1P}
\hat{H}_\st{LF}+\hat{H}_\st{SO}+\hat{H}_\st{B}=
\sum_{i j} \left( h^{i j}_\st{LF} + h^{i j}_\st{SO} + h^{i j}_\st{B} \right) c^{\dagger}_{i}  c_{j}
%\begin{equation}\label{HLF}
%\hat{H}_\st{LF}= \sum_{ \sigma m m'}   H_{m m'}   c^{\dagger}_{m \sigma}  c_{m' \sigma}
\end{equation}
and which are quadratic in the Fermi creation $c^{\dagger}_{i}$ and annihilation $c_{j}$ operators with the combined indices
$i=\{ m , \sigma \}$ and $j=\{ m' , \sigma' \}$. The LF part
\begin{equation}\label{HLF}
h^{i j}_\st{LF}=H_{m m'} \delta_{\sigma \sigma'}
\end{equation}
can be expressed in terms of Steven's operators. That part is specific to each of the following examples and will be detailed there. Finally,
the spin-orbit coupling and the  Zeemann term
\begin{equation}
\label{eq:socZ}
h^{i j}_\st{SO} + h^{i j}_\st{B}= \zeta \left\langle i \left|    \hat{\vec{s}} \, \hat{\vec{l}}  \; \right|    j \right\rangle + 
\mu_B \vec{B} \left\langle i  \left|   \hat{\vec{l}}+ g_s \hat{\vec{s}}  \; \right|  j \right\rangle 
 %\label{eq:socZ}
%\hat{H}_\st{SO}+\hat{H}_\st{B}= \zeta \sum_{ i }   \hat{\vec{s_i}} \hat{\vec{l_i}}   
%+ \sum_{ i }   \mu_B \vec{B}(\hat{\vec{l_i}}+ g_s\hat{\vec{s_i}}) \label{eq:socZ}
\end{equation}
are expressed as matrix elements of the one-particle spin and orbital momentum operators, 
and where $g_s = 2.0023$ is the free electron gyromagnetic ratio.
Please remind, that the parameters of the one-particle Hamiltonian are obtained by DFT calculations, but
the Coulomb parameters from free ions.

To calculate the optical and X-ray spectra the dipole transition probabilities are calculated in the 
ELISA code as it was already published for XAS and XMCD \cite{Giovanelli14}. 
To determine the RIXS intensity we use the formula
\begin{equation}\label{RIXS1}
I(\omega)=\sum_f \delta (\omega_{in}-\omega_{out}-(E_f-E_i)) \lvert A_f \lvert^2
\end{equation}
where $\omega=\omega_{in} - \omega_{out}$ is the energy transfer, the indices $i$ and $f$ denote initial and final states, 
respectively, and with the scattering amplitude

\begin{equation}\label{RIXS2}
A_f=A_f(\omega_{in}) = \sum_m \frac{ 
\langle f \lvert \vec{E}_{in} \cdot \vec{r} \lvert m \rangle 
\langle m \lvert \vec{E}_{out} \cdot \vec{r} \lvert 0 \rangle }
{\omega_{in}-(E_m-E_i)+i \Gamma }
\end{equation}
where we sum over all intermediate states $m$. 
The optical absorption spectra are calculated 
by using the approach of Sugano and 
Tanabe \cite{Sugano70} where the $d$-$d$ transitions between two states $a$ and $b$ become possible by 
combining a parity changing perturbation $V_\st{odd}$ with the dipole operator $\vec{P}=q\cdot \vec{r}$ to
give the transition probability by
\begin{equation}
\label{o1}
W=\left | \frac{2}{\Delta E} \langle a  | V_\st{odd} \vec{P}  | b \rangle     \right |^2 \; , 
\end{equation}
where $\Delta E$ is the energy difference between the given configuration with incomplete $d$ shell 
and the first excited configuration 
with odd parity. We applied this option of ELISA to the case of ZnO:Co and specify the parity breaking 
perturbation later on.

\section{Results}

\subsection{Nickelate La$_2$NiO$_4$}
That compound is a 2D antiferromagnet and an interesting reference material for
2D cuprates since it shares the same crystal structure. But in difference to
the famous La$_2$CuO$_4$ \cite{Bednorz1986,Plakida_book} the local spin
is not $S=1/2$ but $S=1$, instead.  The presence of a 3$d^8$ configuration in a
tetragonal environment allows for interesting multiplet effects and we will use
it here to demonstrate our method. The main focus is to 
calculate, without adjustable parameter, the RIXS
%simulate {\color{blue}(without adjustable parameters)} the RIXS
spectra of Ni$^{2+}$ in La$_2$NiO$_4$ 
which were recently measured
\cite{Fabbris2017} and to compare our method with quantum chemical wave
function methods \cite{Xu2019}. 
%This report contains the simulation results of the RIXS spectra of Ni$^{2+}$
%ions in La$_2$NiO$_4$. The experimental spectra [1]  were taken from the
%thesis of L. Xu [2]. The simulation was performed with a ligand field model
%which is able to describe 
To simulate the RIXS spectra, we use two configurations in the ELISA code,
namely the $3d^82p^6$  ground-state configuration of initial and final states as well as the
excited-state configuration with one core-hole $3d^92p^5$ for the intermediate states of the
RIXS scattering process. The Hamiltonian contains all the Coulomb interactions
within each of the $3d$ and the $2p$ shells as well as between both shells. The
influence of the 6 surrounding oxygen ligands (tetragonal ligand field) is
taken into account by a ligand field Hamiltonian written in terms of Steven's
operators. 

%The Hamiltonian is diagonalized exactly using the ELISA program written by A.
%Savoyant. This program was used before to simulate the XAS spectra of Mn-TCNB
%(Mn-tetracyanobenzene) [3]. Recently, the RIXS simulation was incorporated and
%that is used in the present report.

\subsubsection{Free ion}

The above Hamiltonian {\color{blue}(\ref{H})} without the ligand-field $H_{\st{LF}}$ and Zeeman $H_{\st{B}}$ parts,  describes already very well the energy level scheme of a 
free 
Ni$^{2+}$ ion (NIST data base \cite{NIST_ASD}). That fixes the intra-$d$ Coulomb 
parameters to $F^{(2)}=9.8$ eV and $F^{(4)}=6.1$  eV,  as well as a small spin-orbit 
coupling in the $3d$ shell
of $\zeta=80$  meV. The Slater parameter $F^{(0)}$ is not 
relevant since initial and final 
states have the same number of $d$-electrons. In Table \ref{Tab1} we show the very good agreement between 
experimental and calculated levels.

\begin{table}[h!]
%\begin{center}
\caption
{Comparison of the experimental free ion levels of Ni$^{2+}$ from the NIST data set
\cite{NIST_ASD} 
with the ELISA calculation using the parameters given in the text.} 
\label{Tab1}

%\begin{ruledtabular}  
%\begin{tabular}{lrrr}
%  & This work &  QC & RIXS\\
%\hline
%$\Delta _{x^2-y^2}$ & 0    & 0 & 0 \\
%$\Delta _{xy}$      & 1.67 & 1.36 & 1.38 \\
%$\Delta _{zx,zy}$   & 2.04 & 2.02 & 1.69  \\
%$\Delta _{z^2}$     & 2.24 & 2.38 & 2.39
%\hline
%\end{tabular}
%\end{ruledtabular}
\vspace*{0.3cm}
\begin{tabular}{| c | c | c  |} 
\hline
\hline
%$\mathbf{U_{eff}}$ \textbf{(eV)}   & $\mathbf{E_g$} \textbf{spin up (eV)} &  $ \mathbf{E_g}$ \textbf{spin down (eV)} \\
%\begin{center}
Notation   & ELISA  (eV) & NIST (eV)  \\  
%   & (eV) & (eV) \\
\hline
${}^3 F_4$ & 0.00 & 0.00 \\
${}^3 F_3$   &  0.162 & 0.168 \\
${}^3 F_2$  &  0.272 & 0.281 \\
${}^1 D_2$   &  1.715 & 1.739 \\
${}^3 P_2$   &  2.085 & 2.066 \\
${}^3 P_1$   &  2.125 & 2.105 \\
${}^3 P_0$   & 2.156 & 2.136 \\
${}^1 G_4$   & 2.663 & 2.865 \\
${}^1 S_0$   & 6.399 & 6.513 \\
\hline
\hline
%\end{center}
\end{tabular}
%\end{ruledtabular}
%\end{center}
\end{table}

\subsubsection{Wannier fit}
%\oj{
Stoichiometric \lanio\ features an involved phase diagram with an orthorhombic
and two tetragonal phases~\cite{RodriguezCarvajal1991}. Since the RIXS
measurements were carried out at 20\,K~\cite{Fabbris2017}, at which the
tetragonal $P4_2/ncm$ structure is dominant~\cite{RodriguezCarvajal1991}, we choose this
structure for further analysis. Note that despite the tetragonal symmetry, the
local $z$ axes of neighboring elongated NiO$_6$ octahedra are neither mutually
parallel, nor coincide with the crystallographic $c$ direction. Moreover, the
four short Ni--O bonds split into two pairs of equivalent bonds with the
lengths of 1.944 and 1.954\,\r{A}, respectively. In the following, we refer
to this structure as the experimental structure, to distinguish it from the
simplified (fictitious) $I4/mmm$ structure. The lattice constants for the
latter are chosen so that the constituent NiO$_6$ octahedra have a tetragonal
point symmetry and the Ni--O bond lengths are as close as possible to those of the
experimental structure. Atomic coordinates of both structures are provided in
Table~\ref{tab:lanio_str}.

\begin{table}[tb]
\caption{\label{tab:lanio_str} Atomic coordinates of the experimental (sp.\
gr.\ $P4_2/ncm$ (138); $a$ = 5.4995\,\r{A}, $c$ = 12.5052\,\r{A}) and
simplified (sp.\ gr.\ $I4/mmm$ (139); $a$ = 3.8983\,\r{A}, $c$ =
12.58047\,\r{A}) \lanio\ structures.
}
\begin{ruledtabular}
\begin{tabular}{rcddd}
\multicolumn{5}{c}{experimental \lanio\ structure (from Ref.~\cite{RodriguezCarvajal1991})} \\
\multicolumn{5}{c}{} \\
\multicolumn{1}{c}{site} & \multicolumn{1}{c}{Wyckoff position} & \multicolumn{1}{c}{$x/a$} & \multicolumn{1}{c}{$y/b$} & \multicolumn{1}{c}{$z/c$} \\
Ni & $4d$ &   0            &       0       &   0       \\
La & $8i$ &  -0.0072       &      -0.0072  &   0.3639  \\
 O & $4e$ &  \nicefrac14   &  \nicefrac14  &  -0.0155  \\
 O & $4a$ &  \nicefrac14   &  \nicefrac34  &   0       \\
 O & $8i$ &   0.0314       &       0.0314  &   0.1771  \\ \hline
\multicolumn{5}{c}{simplified (fictitious) \lanio\ structure} \\
\multicolumn{5}{c}{} \\
\multicolumn{1}{c}{site} & \multicolumn{1}{c}{Wyckoff position} & \multicolumn{1}{c}{$x/a$} & \multicolumn{1}{c}{$y/b$} & \multicolumn{1}{c}{$z/c$} \\
Ni & $2a$ &          0 &             0 &   0       \\
La & $4e$ &          0 &             0 &   0.6361  \\
 O & $4c$ &          0 &   \nicefrac12 &   0       \\
 O & $4e$ &          0 &             0 &   0.8229  \\
\end{tabular}
\end{ruledtabular}
\end{table}

\begin{figure}[tb]
  \includegraphics[width=8.6cm]{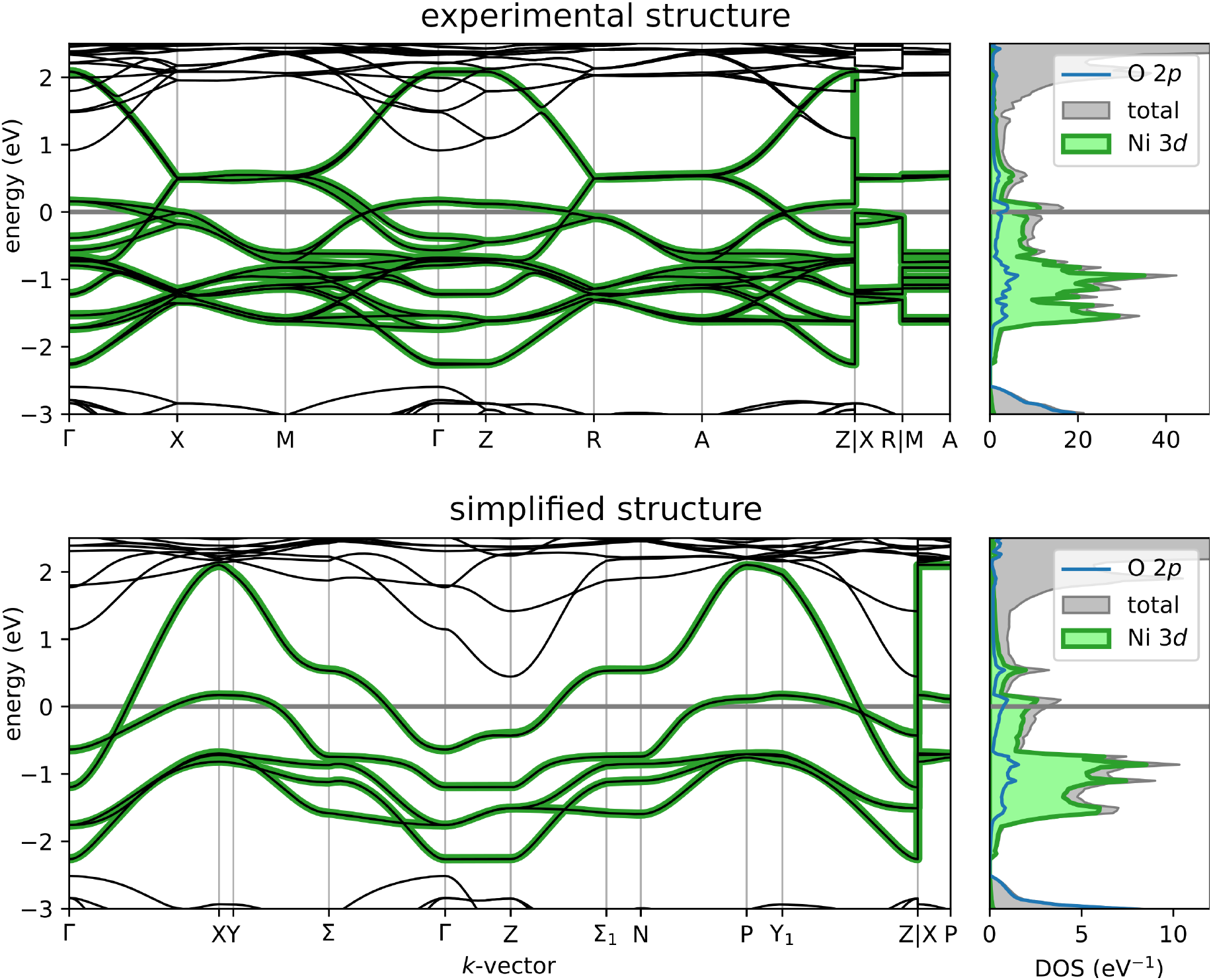}
  \caption{\label{fig:lanio}GGA band structures (thin black lines) and
Fourier-transformed Wannier projections onto Ni $3d$ states (thick green lines)
for the experimental (top) and simplified (bottom) structure of \lanio. Note
that the experimental structure features two Ni sites per cell giving rise to
ten $3d$ bands, while the simplified structure has only one Ni site per
primitive cell, and hence only five bands. The right panels show the densities
of states (DOS): total (gray-shaded), Ni $3d$ (green-shaded), and O $2p$
(blue). For the notation of $k$-vectors, see Ref.~\cite{Aroyo2014}. The Fermi
level is at zero energy.
  }
\end{figure}

We start with scalar-relativistic calculations on $18\times18\times8$
($18\times18\times18$) mesh of $k$-points for the experimental (simplified)
structure.  Typical for correlated insulators, the nonmagnetic GGA yields a
metallic solution, signalled by the bands crossing the Fermi level
(Fig.~\ref{fig:lanio}).  Site- and orbital-resolved densities of states reveal
the dominance of Ni $3d$ states, with a significant admixture of O $2p$ states
(Fig.~\ref{fig:lanio}, right panels). To construct a minimal model, we perform
Wannier projections onto all Ni $3d$ states; the local axes that determine the
Wannier orbital basis coincide with the axes of the respective NiO$_6$
octahedron. We choose the energy window between $-2.4$\,eV and 2.2\,eV (2\,eV
for the simplified structure), the width of the Gaussian tails is 0.3\,eV.  In
this way, we obtain a good agreement between the Fourier-transformed Wannier
Hamiltonian and the GGA band structure (Fig.~\ref{fig:lanio}, left panels).

Since we are interested in local processes, we restrict our analysis to local
hopping terms $t_{ii}$ comprising the $5\times{}5$ onsite Hamiltonian matrix
$H_{ii}$.  
%{\color{blue} (To be consistent with formulation of Sec. II, it would be preferable to write the one-electron matrix element $H_{ii}$ with lower case $h_{ii}$, as in Eq.\ref{HLF}.   
%RH: I repeat my opinion that I do not find that to be a good idea
%)}. 
All Ni sites in either structure are equivalent, hence it is
sufficient to consider a single site $i$.  The tetragonal crystal-field
parameters $D_t$ and $D_s$ as well as the cubic crystal-field splitting
$10D_q$ are obtained by solving the linear problem
\begin{equation}
\label{eq:lanio_cf}
\left(\begin{array}{rrrr}
    1 & -4 & -1 &  2 \\
    1 & -4 &  4 & -1 \\
    1 &  6 & -6 & -2 \\
    1 &  6 & -1 &  2 \\
\end{array}\right)
\left(\begin{array}{c}
    \varepsilon_0 \\
    D_q \\
    D_t \\
    D_s \\
\end{array}\right) =
\left(\begin{array}{c}
    \langle{d_{xy}}|H_{ii}|{d_{xy}}\rangle \\
    \langle{d_{yz/xz}|H_{ii}}|{d_{yz/xz}}\rangle \\
    \langle{d_{3z^2-r^2}}|H_{ii}|{d_{3z^2-r^2}}\rangle \\
    \langle{d_{x^2-y^2}}|H_{ii}|{d_{x^2-y^2}}\rangle \\
\end{array}\right),
\end{equation}
where the coefficients of the matrix are adopted from
Ref.~\cite{Ballhausen1962}. The vector in the right-hand side of the equation
comprises diagonal elements of $H_{ii}$ listed in Table~\ref{tab:lanio_cf}
together with the resulting crystal-field parameters for both structures.

\begin{table*}[tb]
\caption{\label{tab:lanio_cf} Diagonal matrix elements of $H_{ii}$ obtained using Wannier projections onto Ni $3d$ states for scalar-relativistic (no SOC) and full-relativistic (with SOC) GGA calculations of the experimental as well as the simplified structure of \lanio. The cubic ($10D_q$) and tetragonal ($D_t$, $D_s$) crystal-field parameters are calculated via Eq.~(\ref{eq:lanio_cf}). All values are in meV.}
\begin{ruledtabular}
\begin{tabular}{cddp{.005\textwidth}dd}
& \multicolumn{2}{c}{experimental structure} & & \multicolumn{2}{c}{simplified structure} \\
& \multicolumn{1}{c}{no SOC} & \multicolumn{1}{c}{with SOC} & & \multicolumn{1}{c}{no SOC} & \multicolumn{1}{c}{with SOC} \\ \hline
$\langle{d_{xy}}|H_{ii}|{d_{xy}}\rangle$             & -1246.7 & -1250.1 & & -1250.6 & -1254.9 \\
$\langle{d_{yz}|H_{ii}}|{d_{yz}}\rangle$             & -1190.4 & -1193.7 & & -1197.1 & -1200.4 \\
$\langle{d_{3z^2-r^2}}|H_{ii}|{d_{3z^2-r^2}}\rangle$ & -1189.7 & -1195.8 & & -1197.1 & -1200.4 \\
$\langle{d_{xz}|H_{ii}}|{d_{xz}}\rangle$             &  -126.8 &  -130.6 & &  -125.1 &  -131.2 \\
$\langle{d_{x^2-y^2}}|H_{ii}|{d_{x^2-y^2}}\rangle$   &   336.5 &   333.4 & &   329.8 &   329.9 \\ \hline
$10D_q$                                              &  1583.2 &  1583.5 & &  1580.4 &  1584.9 \\
$D_t$                                                &    46.2 &    46.1 & &    45.1 &    45.8 \\
$D_s$                                                &    58.1 &    58.4 & &    57.3 &    58.1 \\
\end{tabular} 
\end{ruledtabular}
\end{table*}

Another key parameter that we can extract from DFT calculations is the  3$d$-shell
spin-orbit coupling constant $\zeta$. To this end, we perform full-relativistic
nonmagnetic GGA calculations, followed by a wannierization 
(see Appendix \ref{App:Wsoc} for details). We focus again on
the onsite Hamiltonian matrix $H_{ii}$, which is now a $10\times{}10$ matrix
due to the presence of two spin channels. Let us start with the simplified
structure. Here, the $H_{ii}$ in Table \ref{tab:Hii_lanio_simpl} has precisely
the same form as the sum of the diagonal crystal-field and the atomic
spin-orbit (Table~\ref{tab:SOC}) contributions. Since the latter has only one
free parameter, the spin-orbit coupling constant $\zeta$, we can estimate it by
comparing the respective offdiagonal matrix elements of $H_{ii}$ with that of
Table~\ref{tab:SOC}.  Since the latter imply the local cubic symmetry which is
higher than the tetragonal symmetry of our simplified structure, different
matrix elements give slightly different values of $\zeta$. By averaging over
all matrix elements, we obtain $\zeta=73\pm3$\,meV.

Now we turn to the more complicated case of the experimental structure. Here,
the matrix form of $H_{ii}$ in 
Table \ref{tab:Hii_lanio_exp}
%Eq.~(\ref{eq:Hii_lanio_exp}) 
substantially
differs from that of the simplified structure. The reason is twofold. First,
the local symmetry of Ni sites is orthorhombic. As a result, some off-diagonal
matrix elements in the crystal-field Hamiltonian become nonzero. Second, the
$x$ axis of the coordinate system utilized by FPLO does not coincide with the
local $x$ axis of NiO$_6$ octahedra; a $\frac34{}\pi$ rotation would put it in
place. While the current version of FPLO does not allow the user to choose the
direction of the $x$ axis, we can rotate the basis of the matrix describing the
atomic SOC such that has the same form as our $H_{ii}$. The technical details
of this transformation are described in the Appendix~\ref{App:LSRot}. In this
way, we obtain $\zeta=74\pm2$\,meV, which is in excellent agreement with the value
obtained for the simplified structure.

\subsubsection{Multiplet spectra}

In the given case of tetragonal symmetry the ligand-field part of the Hamiltonian writes:
\begin{equation}
%H_\st{LF}= B_{20} O^0_2 + B_{40} O^0_4 + B_{44} O^4_4 
\hat H_{\mathrm{LF,t}} = B_{20}\hat{O}_{2}^{0}+B_{40}\hat{O}_{4}^{0}
 +B_{44}\hat{O}_{4}^{4} \label{tetrStev}
\end{equation}
%{\color{blue} (Same remark as above: $H_{LF}$ must be wrtitten with lower case. Also, I would remove the index t, which 
%complicate the notation for none advantage)}.\\
where the explicit expressions of the Steven's 
operators are repeated in the Appendix \ref{App:Stev}.
Sometimes, the three crystal field parameters are expressed differently by 
\begin{equation}
D_q=\frac{12}{5} B_{44} \, , \; D_s=3 B_{20}  \, , \; D_t=\frac{12}{5} B_{44} - 12 B_{40} \; .
\end{equation}
To calculate the multiplet spectrum (see Table \ref{Tab2}) we use the parameters $10D_q=1.5835$ eV, $D_s=0.0584$ eV, 
and $D_t=0.0461$ eV; corresponding to  $B_{44}=65.98$ meV, $B_{40}=9.35$ meV, 
and $B_{20}=19.47$ meV; as obtained from a Wannier fit to the 
bandstructure of the nonmagnetic DFT solution of La$_2$NiO$_4$ in the realistic tetragonal structure.
In Table \ref{Tab2} we compare the Wannier parameter spectrum with the experimental one as obtained by RIXS, as well 
as with a spectrum obtained by a quantum chemical wave function method using 
the complete-active-space self-consistent-field (CASSCF) and
multireference configuration-interaction (MRCI) techniques,
finding a good agreement. 

\begin{table}[h!]
\begin{center}
\caption{
Comparison of multiplet energies as obtained from the ELISA program with LF parameters from a Wannier fit 
with quantum chemical calculations (QC) \cite{Xu2019} and experimental RIXS data \cite{Fabbris2017}.  
%quantum chemistry (Xu thesis [2]) and with Wannier LF parameters. The last column compares with the experimental spectrum (see supplemental
%material in [1])
(Spin-orbit coupling neglected for simplicity).} 
\label{Tab2}

\begin{tabular}{| c | c | c | c  |} 
\hline
\hline
%$\mathbf{U_{eff}}$ \textbf{(eV)}   & $\mathbf{E_g$} \textbf{spin up (eV)} &  $ \mathbf{E_g}$ \textbf{spin down (eV)} \\
Notation   & Quantum chemistry \cite{Xu2019}  & This work & Experiment \cite{Fabbris2017} \\  
   & (eV) & (eV) & (eV) \\
\hline

${}^3 B_{1g}$ & 0.00 & 0.00 & 0.00 \\
${}^3 E_{g}$   &  0.74 &1.15 & 1.06 \\
${}^3 B_{2g}$  &  1.30 & 1.58 & 1.61 \\
${}^3 A_{2g}$   &  1.46 & 2.04 & 1.61 \\
${}^1 A_{1g}$   &  1.66 & 1.81 & 1.61 \\
${}^3 E_{g}$   &  1.73 & 2.24 & 2.29 \\
${}^1 B_{1g}$   & 1.92 & 1.94 & 2.29 \\
${}^1 E_{g}$   & 2.59 & 3.03 & 2.93 \\

\hline
\hline
\end{tabular}
\end{center}
\end{table}

\subsubsection{RIXS}

To determine the RIXS intensity we use the Eqns.\ {\color{blue}(\ref{RIXS1},\ref{RIXS2})}.
For the Ni $L_3$-edge RIXS measurements, 
the intermediate states are  those of the $2p^53d^9$ excited configuration, made of two open shells. 
They are separated by 855 eV from the ground configuration containing the initial and final states.
The ELISA program can treat all interactions within and in between the two configurations. 
In addition to the three Coulomb 
parameters of the $3d$ shell we use the Coulomb interaction of the $2p$ shell $F^{(2)}_{2p}=6$ eV,   
%{\color{blue}($F_{pp}^2$: this parameter can be set to zero, as the p-shell is either full or occupied by one hole. Maybe better not to mention this parameter: the %situation is sufficiently complicated!)} and, {\color{blue} for excited configuration,} the 
inter-shell Coulomb 
parameters $F^{(2)}_{pd}=5.81$ eV, 
%{\color{blue}($F_{pd}^2$)}, 
$G^{(1)}_{pd}=4.32$ eV,  
%{\color{blue}($G_{pd}^1$)}
and $G^{(3)}_{pd}=2.46$ eV. 
%{\color{blue}($G_{pd}^3$)}  
These 
parameters are obtained from a 
Hartree-Fock solution \cite{Groot2005,Groot2010} of the free ion, but they are not critical for the RIXS spectra.

\begin{figure}[h!]
\centering
\includegraphics[%draft, 
width=\columnwidth]{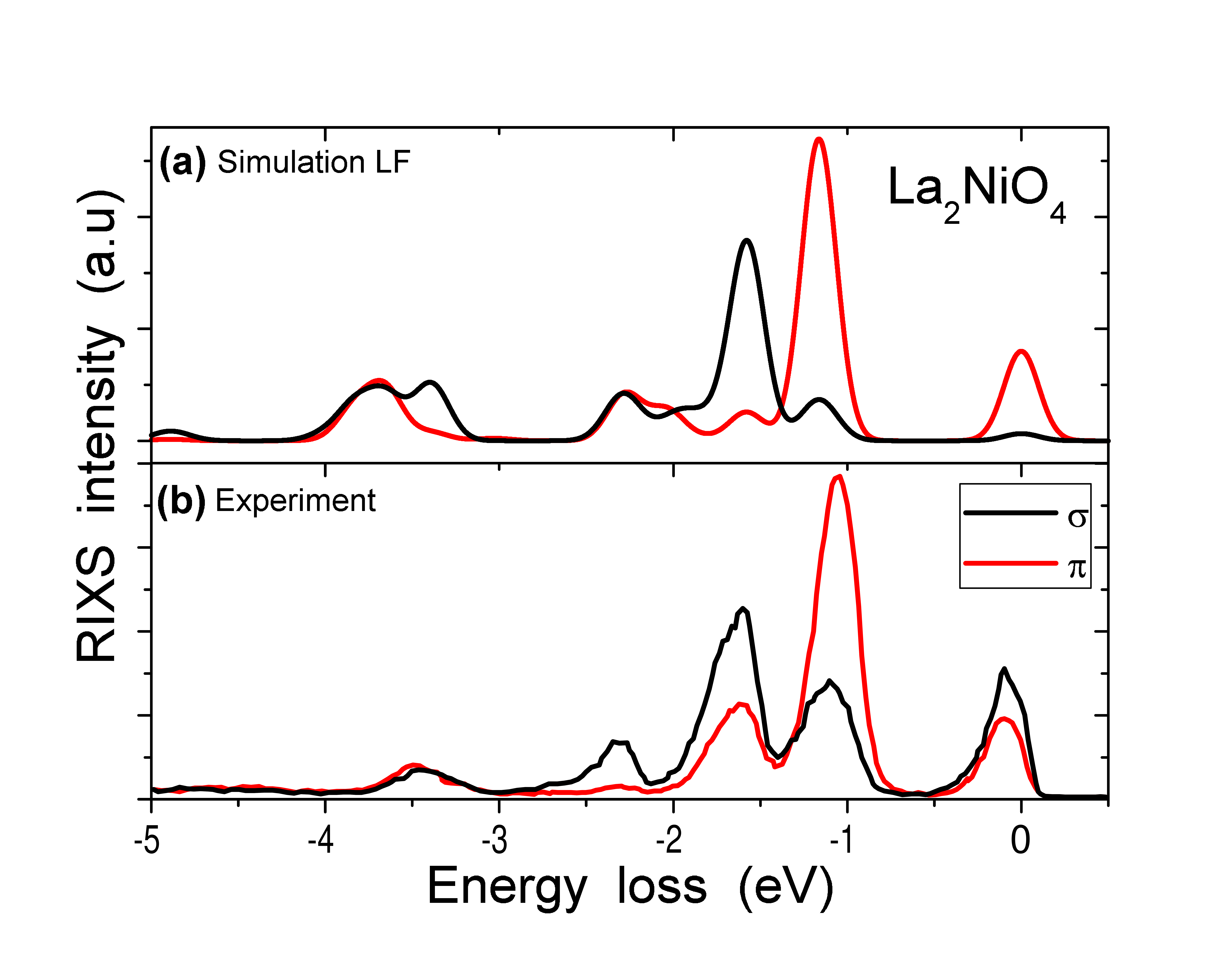}
\caption{ 
Comparison of experimental \cite{Fabbris2017}  and theoretical RIXS spectra
for $\pi$- and $\sigma$-polarization and an income angle of $\Theta=20 \degree$ 
(The Gaussian broadening of the theoretical curve was set to 0.24 eV; see text for more details).}
\label{RIXSNi}
\end{figure}

The inclusion of the core-hole configuration does not influence the energy level 
scheme of the $3d^8$ configuration. It is just 
needed to calculate the matrix elements of the scattering amplitude. The energy of the incoming 
X-rays $\omega_{in}$ is chosen such 
that we integrate over all states of the $J=3/2$ manifold, i.e. we restrict ourselves to the $L_3$-edge. 
The SO coupling splits the groundstate triplet $^3B_{1g}$ (see Table \ref{Tab2})  into a lower singlet ($B_{2g}$) and an upper doublet ($E_g$)
which are separated by 1.1 meV in our approach, and by 2.1 meV in quantum chemistry \cite{Xu2019}. 
The directions of the electric field for incoming and outgoing X-rays are dictated by the experimental geometry (see Fig. 8.6 c 
in \cite{Xu2019}). The incoming X-rays are linearly polarized, either perpendicular ($\sigma$-polarization) to the scattering plane
or within the scattering plane ($\pi$-polarization).  
The c-axis is directed perpendicular to the surface and the incoming X-ray has an angle of $\Theta$ with the surface. The angle 
between incoming and outgoing X-rays is fixed to $\alpha=50 \degree$. 
Accordingly, the direction of the electric field for the incoming beam with $\pi$- and $\sigma$-polarization is given by:
\begin{equation}
E^{\pi}_{in}=\sin \Theta \vec{e}_x + \cos \Theta \vec{e}_z \qquad  E^{\sigma}_{in}=\vec{e}_y
\; .
\end{equation}
For both cases, we sum the intensities for both possible polarizations of the outgoing beam, i.e.
\begin{equation}
E^{(1)}_{out}=\sin (\Theta+\alpha) \vec{e}_x + \cos (\Theta+ \alpha) \vec{e}_z \qquad  E^{(2)}_{out}=\vec{e}_y
\; .
\end{equation}
In Fig.\ \ref{RIXSNi} we compare the calculated spectra with experiment \cite{Fabbris2017}. 
We choose an income angle of
$\Theta=20 \degree$ since there exist detailed data showing a pronounced dichroism for that income angle. We  
also calculated the RIXS spectra for other income angles (not shown) and find generally
a good agreement with experiment as concerns the main peak positions, its relative weight, and angle dependence. 
The experimental low-energy peak at about -0.1 eV energy loss is shifted from the zero of energy due to a local 
exchange field of the neighboring spins in the antiferromagnetic structure which is possible for an income angle
of $\Theta=20 \degree$
and is situated at zero energy in the theoretical curve.
%Without exchange field (see Fig.\ \ref{RIXSNi} a) that peak would be situated at zero and 
%the inclusion of an exchange field of $B_{ex}=0.001$ eV improves the peak position, but greatly reduces its intensity for both polarizations, 
%and influences also slightly some
%of the other peaks.

\subsection{High-$T_c$ parent compound $\mathrm{CaCuO_2}$}

\begin{figure}[htb] %%%%%%%%%%%%%%%%%%%%%%%%%%%%%%%%%%%%%%%%%%%%%%%%%%%%%%%%%%%%%%%%
\includegraphics[%draft, 
%width=0.95\columnwidth]{fig3}
width=9.2cm]{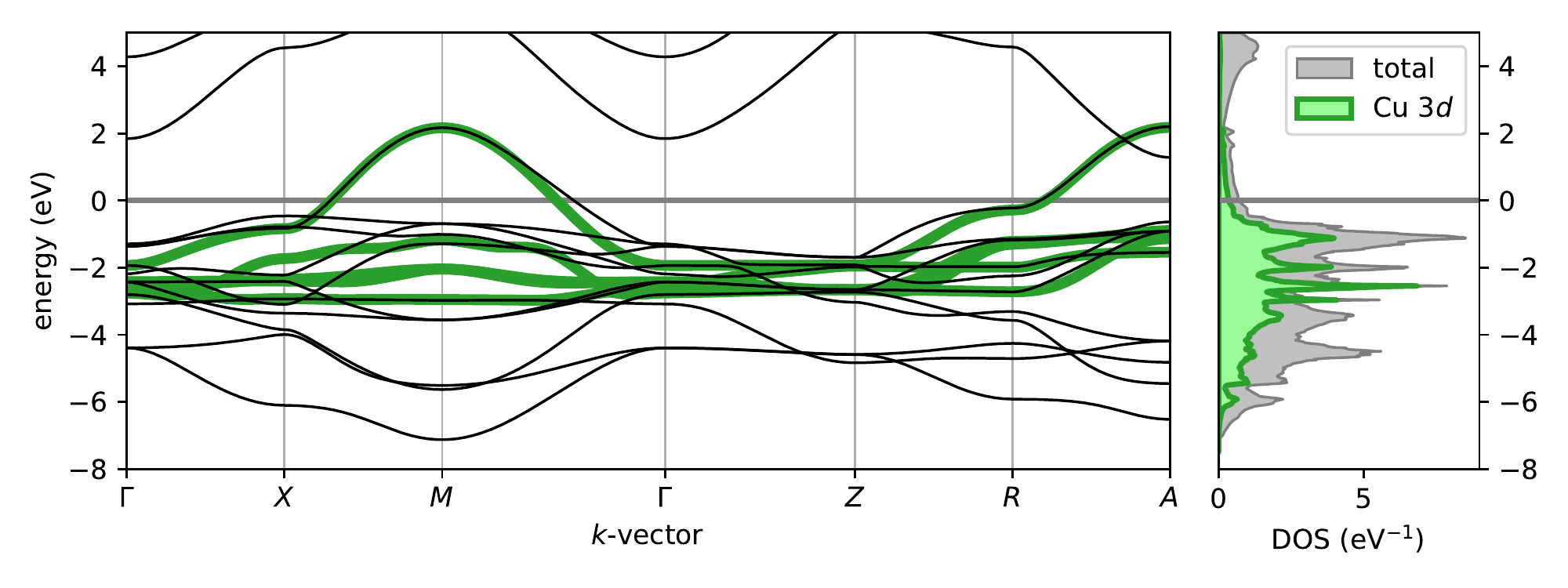}
\caption{\label{cco}
Band structure (thin black lines) and Wannier projection (thick green lines), as well as densities 
of states for the 
%Density of states for the 
high-$T_c$ parent compound $\mathrm{CaCuO_2}$. The Fermi level $\varepsilon_F$ is set 
to zero.
}
\end{figure} %%%%%%%%%%%%%%%%%%%%%%%%%%%%%%%%%%%%%%%%%%%%%%%%%%%%%%%%%%%%%%%%%%%%%
Soon after the discovery of high-$T_c$ superconductivity in doped 
$\mathrm{La_2CuO_4}$ \cite{Bednorz1986}, P.W. Anderson \cite{Anderson1987} 
realized that 
strong correlations in the copper $3d$ shell and the quasi-two-dimensional layered
structure of this and related copper oxide compounds are substantial 
for the understanding of their physics \cite{Plakida_book}. The common
structural unit for the rich family of cuprate high-$T_c$ superconductors
is the CuO$_2$ plane that is built from corner-shared CuO$_4$ plaquettes. 
In each plaquette, the Cu$^{2+}$ ion is surrounded by a square of four 
oxygen ligands. The LF strongly splits the Cu $3d$ levels so that
only the $d_{x^2-y^2}$ orbital hybridized with oxygen $p_{\sigma}$ orbitals
contributes to the states in the vicinity of Fermi energy and defines the
low energy physics of the  high-$T_c$ cuprates. The so-called parent (undoped)
compounds contain exactly one hole per CuO$_4$ plaquette. 
The strong correlations suppress charge fluctuations. The low-energy physics 
is defined by spin degrees of freedom. The parent compounds are charge-transfer
insulators described by the antiferromagnetic Heisenberg Hamiltonan. The localized holes  
reside mainly on the copper $d_{x^2-y^2}$ orbitals. A doping introduces
extra holes or electrons into the CuO$_2$ planes. With the increase of 
carrier concentration, the system becomes metallic and superconducting. 

The $\mathrm{CaCuO_2}$ is a so-called infinite-layer compound, 
with equal distances between the CuO$_2$ planes, which are separated only
by Ca cations. Under hole doping, the $T_c$ of superconductivity
reaches 110~K \cite{Azuma1992}. Recently, the discovery of superconductivity in hole doped
NdNiO$_2$, which is almost isoelectronic to CaCuO$_2$, renewed the interest to 
the electronic structure of this compound 
\cite{Botana2020,Karp2020}.

This undoped layered cuprate has tetragonal space group $P4/mmm$(\#123) with lattice parameters 
$a=b=3.86$~\AA, $c=3.20$~\AA. The band structure and projected densities 
of states of the nonmagnetic GGA solution are depicted in Fig.\ \ref{cco}. The figure shows that 
the antibonding mixture of oxygen $p$- and copper $d$-states (corresponding to five $d$-like 
bands in the band structure) lies within 
the energy window $-3 < \varepsilon - \varepsilon _F< 2.5$~eV (see Appendix \ref{App:chewin} 
for the choice of the energy window). Within this window, we have 
constructed Wannier functions  
having cubic harmonic symmetry ($d_{xy}$, $d_{xz}$, $d_{zy}$, $d_{x^2-y^2}$, 
and  $d_{z^2}$). 
The $d_{x^2-y^2}$ orbital has the highest energy.
In the ground state, a hole occupies this orbital and
the $d$-$d$ excitations correspond to electron transitions from low lying 
$d$-levels to the empty $d_{x^2-y^2}$ orbital.
Thus, the $d$-$d$ excitation energies can be calculated 
by  
the onsite energy differences $\Delta _i=E_{x^2-y^2}-E_i$ ($E_{x^2-y^2}= \varepsilon _F-0.094$~eV) for 
the Wannier functions. 
These energy differences 
are given in
the second column of Table \ref{tab:cco} and are in good agreement both 
with sophisticated quantum chemical calculations \cite{Hozoi2011} and
experimental values from RIXS measurements \cite{Sala2011}. Following 
Ref. \cite{Hozoi2011}, the RIXS energies are corrected by the magnetic energy 
value $\Delta E_{\mathrm{magn}}=0.26 $~eV.
%
%-3 < E < 2.5
%xy  1.672
%xz,yz  2.038
%z2  2.236
\begin{table}[tbp]
\caption{\label{tab:cco} Results for the calculated Cu $d$-level splitting (in eV)  of  
$\mathrm{CaCuO_2}$ compared with quantum chemical calculations \cite{Hozoi2011}
and RIXS \cite{Sala2011}. Magnetic energy $\Delta E_{\mathrm{magn}}=0.26 $~eV
is substracted from each of the RIXS values \cite{Hozoi2011}.
}
\begin{ruledtabular}  
\begin{tabular}{lrrr}
  & This work &  QC & RIXS\\
\hline
$\Delta _{x^2-y^2}$ & 0    & 0 & 0 \\
$\Delta _{xy}$      & 1.67 & 1.36 & 1.38 \\
$\Delta _{zx,zy}$   & 2.04 & 2.02 & 1.69  \\
$\Delta _{z^2}$     & 2.24 & 2.38 & 2.39
%\hline
\end{tabular}
\end{ruledtabular}
\end{table}

Then we find the parameters of the Stevens Hamiltonian (\ref{tetrStev}) for $\mathrm{CaCuO_2}$
as
%B44r,B40r, B20r=  69.66666666666667 1.4476190476190443 123.90476190476193
\begin{align}
B_{44} &= \frac{1}{24}\left(E_{x^2-y^2}-E_{xy}\right) \approx 69.7 \text{~meV}, \label{B44tet} \\
B_{40} &= \frac{B_{44}}{5} + \frac{\Delta _t}{105} 
 - \frac{\Delta _g}{140} \approx  1.4  \text{~meV}, \label{B40tet} \\
B_{20} &= \frac{1}{21}\left(\Delta _t + \Delta _g \right) \approx 
 123.9 \text{~meV}. \label{B20tet} 
\end{align}
where we have introduced the notations $\Delta _t \equiv E_{xy}-(E_{yz}+E_{xz})/2$,
$\Delta _g \equiv E_{x^2-y^2}-E_{z^2}$. 

\begin{figure}[h!]
\centering%
\includegraphics[%draft, 
width=1.0 \linewidth]{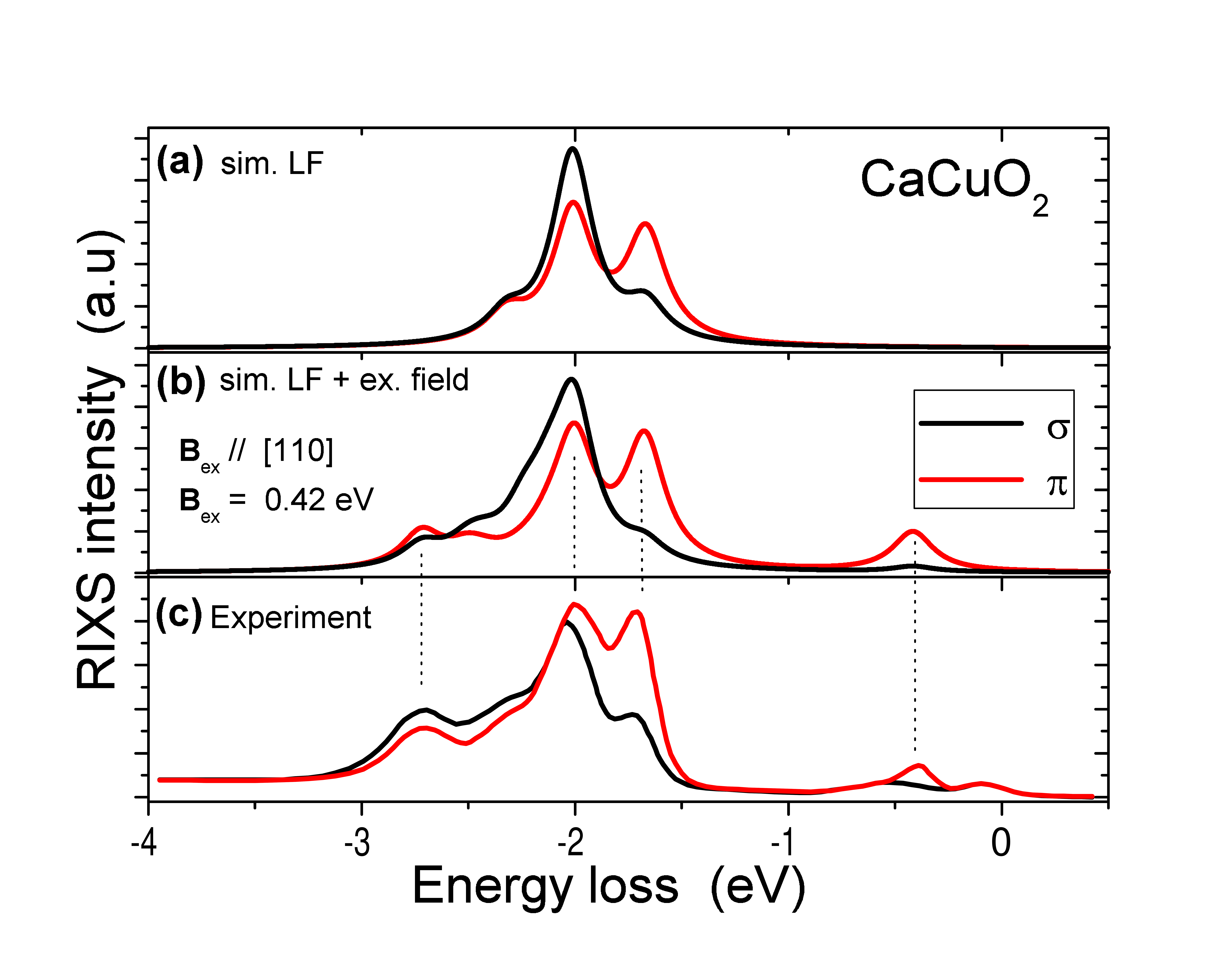}
\caption{ 
Comparison of experimental 
(cf. Fig.2. of Ref. \onlinecite{Sala2011})  and two theoretical (without and with exchange field) RIXS spectra
for $\pi$- and $\sigma$-polarization and an income angle of $\Theta=20 \degree$ 
(The Lorentzian broadening of the theoretical curves was set to 0.23 eV; see text for more details).
\label{RIXSCu}}
\end{figure}

Using the so determined LF parameters we can calculate the multiplet spectrum and the RIXS curves. For a 
$d^9$ configuration, the energy differences in the multiplet spectrum coincide with the energy differences of the
single electron spectrum (see Table \ref{tab:cco}) and they are not at all influenced by the Coulomb parameters.
Correspondingly, the peak positions of the RIXS spectra (see Fig.\ \ref{RIXSCu}) can be interpreted with the help of 
Table \ref{tab:cco}. In the calculated RIXS spectrum without exchange field (Fig.\ \ref{RIXSCu} a) we can easily 
distinguish the excitations for $d_{xy}$, $d_{xz}$/$d_{yz}$, and $d_{z^2}$ orbitals at -1.7, -2.0, and -2.2 eV 
energy loss as single peaks or well developed shoulder. However, the experimental spectrum shows also
a magnetic peak due to magnon excitations at -0.4 eV which we can simulate in our calculation by an 
exchange field of 0.42 eV acting only on the spin moment and being directed towards the crystallographic
[110] direction as in the experimental magnetic structure. It should be noted that an exchange field perpendicular 
to the Cu-O plane does not lead to a magnetic peak 
in accordance with the results of theoretical predictions \cite{Ament2009}. 
Interestingly, the inclusion of an exchange field improves also 
the agreement between experiment and theory at large energy loss and leads to the appearance of peaks
at about -2.6 $\ldots$ -2.7 eV energy loss.

\subsection{Quasi-one-dimensional cuprate $\mathrm{Li_2CuO_2}$}

The compound $\mathrm{Li_2CuO_2}$ belongs to the family of edge-shared cuprates (ESC). 
The states near Fermi energy are provided by CuO$_4$ plaquets that share 
their edges and form CuO$_2$ chains. 
The ESC compounds  represent a particular class of quantum magnets in which the local geometry gives rise 
to competing nearest ferromagnetic or antiferromagnetic exchange coupling $J_1$ and frustrating 
antiferromagnetic next-nearest neighbor $J_2$ coupling. 
The 
one-dimensional spin-\nicefrac{1}{2} $J_1$-$J_2$ Heisenberg model is one prime 
example of frustrated magnetism, where quantum fluctuations can 
alter both ground state and spin correlations\,\cite{Mikeska2004,Drechsler07}. 
Due to its simple structure 
with flat CuO$_2$ chains, the compound $\mathrm{Li_2CuO_2}$ was considered
as a model system for studies of the highly non-trivial magnetism in 
the edge-shared cuprates
\cite{Sapina90,Graaf02,Yushankhay02,Kawamata2004,Malek08,Lorenz09,
Nishimoto11,Monney13,
Nishimoto15,Johnston16,Money16,Kuzian2018}. 
Almost two decades passed from the determination of the crystal and magnetic structures 
of $\mathrm{Li_2CuO_2}$ \cite{Sapina90} to a reliable determination of main magnetic interactions 
within and between the CuO$_2$ chains \cite{Lorenz09,Lorenz11}. This allowed understanding 
the thermodynamics of the compound \cite{Nishimoto11,Nishimoto15,Kuzian2018} and to demonstrate 
how the charge-transfer excitation spectra having the scale of several eV are governed by correlations in 
the spin system with an energy scale $\sim 0.01$~eV. This results in strong temperature dependence of the 
spectra \cite{Malek08,Monney13,Johnston16,Money16}.

\begin{figure}[htb] %%%%%%%%%%%%%%%%%%%%%%%%%%%%%%%%%%%%%%%%%%%%%
%\begin{figure}[h!]
%%%%%%%%%%%%%%
%\includegraphics[draft, trim = 0.0cm 0.0cm 0.0cm 0.0cm, clip=true,width=0.99\columnwidth]{lcob}
\includegraphics[%draft, 
width=1.0\columnwidth]{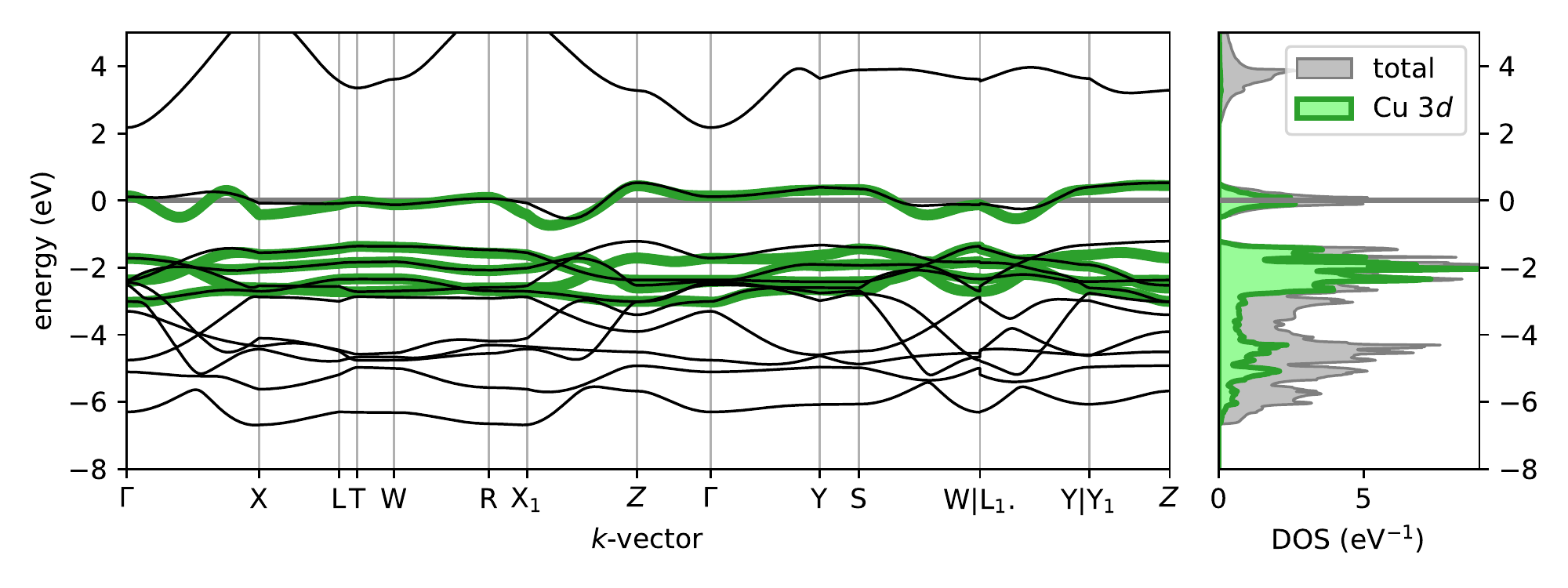}
%width=8.6cm]{fig5}
\caption{\label{lco}
Band structure (thin black lines) and
Wannier projection (thick green lines), as well as 
densities of states %(lower panel) 
for the 
quasi-one-dimensional cuprate compound $\mathrm{Li_2CuO_2}$.
}
\end{figure} %%%%%%%%%%%%%%%%%%%%%%%%%%%%%%%%%%%%%%%%%%%%%%%%%%%%%%%%%%%%%%%%%%%%%

The space group of the crystal structure is $Immm$(\#71),
lattice parameters are $a=3.65445$, $b=2.86022$, $c=9.3774$~\AA.
The CuO$_2$ chains run along $b$ direction in the crystal $bc$ plane.
The copper site has $D_{2h}$($mmm$) point group symmetry. It is convenient to 
take the local coordinate system with the center on the Cu site and the $x$ axis
directed along the crystallographic $b$ direction, axis $y \parallel c$,
and $z\parallel a$. This coordinate system is rotated by $\pi /4$ around 
the $z$-axis compared to the one used for $\mathrm{CaCuO_2}$.
Figure \ref{lco} shows that 
in  $\mathrm{Li_2CuO_2}$ the 
antibonding mixture of oxygen $p$- and copper $d$-states lies within 
an energy window $-3 < \varepsilon - \varepsilon _F< 0.55$~eV. 
%The energies of $d-d$
%excitations compared with quantum chemical and RIXS results are given in
%Table \ref{tab:lco}. 

% 
\begin{table}[tbp]
\caption{\label{tab:lco} Results for the calculated Cu $d$-level splitting (in eV) of  
$\mathrm{Li_2CuO_2}$ compared with quantum chemical calculations
and RIXS \cite{Bogdanov2016}. 
}
\begin{ruledtabular}  
\begin{tabular}{lrrr}
  & This work &  QC & RIXS\\
\hline
$\Delta _{xy}$ & 0    & 0 & 0 \\ 
$\Delta _{zx}$ & 1.82 & 2.02 & 2.1 \\
$\Delta _{zy}$ & 1.96 & 2.09 & 2.1 \\
$\Delta _{x^2-y^2}$ & 1.90 & 1.55 & 1.7 \\
$\Delta _{z^2}$ & 2.28 & 2.85 & 2.6
\end{tabular}
\end{ruledtabular}
\end{table}

In the ground state of $\mathrm{Li_2CuO_2}$, 
a hole occupies the $d_{xy}$ orbital  that lies in the $bc$ plane 
($E_{xy} = \varepsilon _F-0.17$~eV; that orbital would correspond to the $d_{x^2-y^2}$ orbital if we would have 
chosen the same coordinate system as in CaCuO$_2$). 
Because of orthorhombic symmetry, the  $d_{zy}$ and and $d_{zx}$ orbitals
are not degenerated and, 
moreover, a non-diagonal matrix element
$V=\langle d_{x^2-y^2}\left|\hat H_{LF}\right|d_{z^2}\rangle$ is not 
prohibited by symmetry. However, $V\approx 0$ for the 
chosen energy window (see Fig.\ \ref{ewin}b in Appendix \ref{App:chewin}).
The LF Hamiltonian for the Cu site 
in $\mathrm{Li_2CuO_2}$ acquires additional terms compared with 
$\hat H_{\mathrm{LF,t}}$ (\ref{tetrStev})
\begin{align}
\hat H_{\mathrm{LF,o}} &=\hat H_{\mathrm{LF,t}}
+B_{22} \hat{O}_{2}^{2}+B_{42}\hat{O}_{4}^{2}. \label{ortStev} 
\end{align}
The parameters of $\hat H_{\mathrm{LF,t}}$ found from Eqs.\ (\ref{B44tet})-(\ref{B20tet})
are $B_{40} \approx -0.55$, $B_{44} \approx -79.1$, 
$B_{20} \approx 108.1$~meV. The other parameters are
\begin{align}
B_{42} & = -\frac{1}{42}\left(E_{zx}-E_{zy}\right) \approx 3.5 \text{~meV}, \label{B24} \\
B_{22} & = -3B_{42} \approx -10.5 \text{~meV}. \label{B22} 
\end{align}
Table \ref{tab:lco} shows that the energy of $d$-$d$ excitations obtained 
from the Wannier function Hamiltonian are in good agreement with the values 
measured in RIXS experiments and calculated by the elaborated 
quantum chemical approach  \cite{Bogdanov2016}.

A magnetic induction $B_{\alpha}$, $\alpha =x,y,z$ applied along 
a symmetry axis $\alpha $ splits the ground state doublet according 
to the 
effective spin-Hamiltonian
\begin{equation}
\hat{H}_{\mathrm{s}} = g_{\alpha}\mu _B\hat{s}^{\alpha}B_{\alpha}.
\label{eq:Hs}
\end{equation}
The $g$-factors are precisely determined in electron paramagnetic 
resonance experiments. The reported values for $\mathrm{Li_2CuO_2}$ are
$g_a\approx 2.264$, $g_b\approx 2.047$ and $g_c\approx 2.033$ \cite{Kawamata2004}.
The deviation of $g$-factors from the free electron value $g_s$ 
characterizes the 
partial unquenching of the orbital moment  due to the spin-orbit 
interaction. 

To calculate the $g$-factors we need, besides the LF parameters, also the spin-orbit coupling
$\zeta$. For that purpose, we performed a Wannier fit for the full-relativistic but nonmagnetic GGA 
calculation of Li$_2$CuO$_2$ (see Appendix \ref{App:SOC}). In contrast to La$_2$NiO$_4$, we observe a
strong covalency reduction $k$ and large deviations from spherical symmetry. The covalency reduction of the
free ion spin-orbit (SO) coupling $\lambda_0=-103$ meV of Cu$^{2+}$  
\cite{NIST_ASD} 
is defined as $\zeta=-k \lambda_0$. (Here, $\zeta$ has a positive sign since it is defined for single electrons, whereas
$\lambda$ is defined for the total spin and is negative.)
%I always found that one-electron SO coupling constant was positive, at least in free ion. For me, the negative sign arises when effective coupling in  many-electron basis is considered, depending on the shell filling. Can you explain me please?)}
Despite the fact that the covalency reduction is usually applied when the rotational symmetry is preserved,
we feel justified in the present case of Li$_2$CuO$_2$ to take into account only the 
SO matrix elements with 
the $d_{xy}$ ground state where the hole resides predominantly. An average over the three relevant matrix elements gives $k=0.57$ which
can be used in the ELISA program to calculate the $g$-factors. With such a procedure we obtain high precision
of the low-lying part of the multiplet spetrum but accept deviations at higher energies.

In the given special case of a $d^9$ configuration we have a more precise way of determining
the $g$-factors by diagonalization of the 
on-site Hamiltonian matrix in 
the basis of Wannier functions obtained in the full-relativistic calculation.
The calculation (see Appendix \ref{App:SOC} for the details) gives $g_z \approx 2.255$,
$g_x \approx 2.062$, $g_y \approx 2.053$, that is in good agreement with
the experimental values. The calculation thus confirms the strong reduction
of spin-orbit coupling in $\mathrm{Li_2CuO_2}$.

\subsection{Diluted magnetic semiconductor Co doped ZnO}

Co- or Mn-doped ZnO and similar systems were thoroughly studied as diluted magnetic 
semiconductors 
in which room-temperature ferromagnetism was predicted 
due to $p$-doping
\cite{Dietl00}. A ferromagnetic-like behavior at room 
temperature was indeed observed in some samples of ZnO:Co and other 
doped oxides \cite{Dietl10,Ogale10,Simimol15,Dietl15}. The origin of 
this behavior is still under debate \cite{Dietl10,Kuzian16}.

The non-magnetic Zn$^{2+}$ ion of the host lattice is substituted by the Co$^{2+}$
impurity ion having 3 holes in the $3d$-shell and an effective spin $S=\nicefrac{3}{2}$.
One of the fingerprints of the intrinsic magnetism in ZnO:Co is a strong easy-plane magnetic 
anisotropy \cite{Sati06}. The anisotropy is due to the single-ion anisotropy 
of Co$^{2+}$, which is the consequence
of the $d$-level splitting by the trigonal ligand field generated by a tetrahedral 
oxygen surrounding \cite{Kuzian06}.  

\begin{figure}[htb] %%%%%%%%%%%%%%%%%%%%%%%%%%%%%%%%%%%%%%%%%%%%%%%%%%%%%%%%%%%%%%%%
\includegraphics[%draft, 
width=1.0\columnwidth]{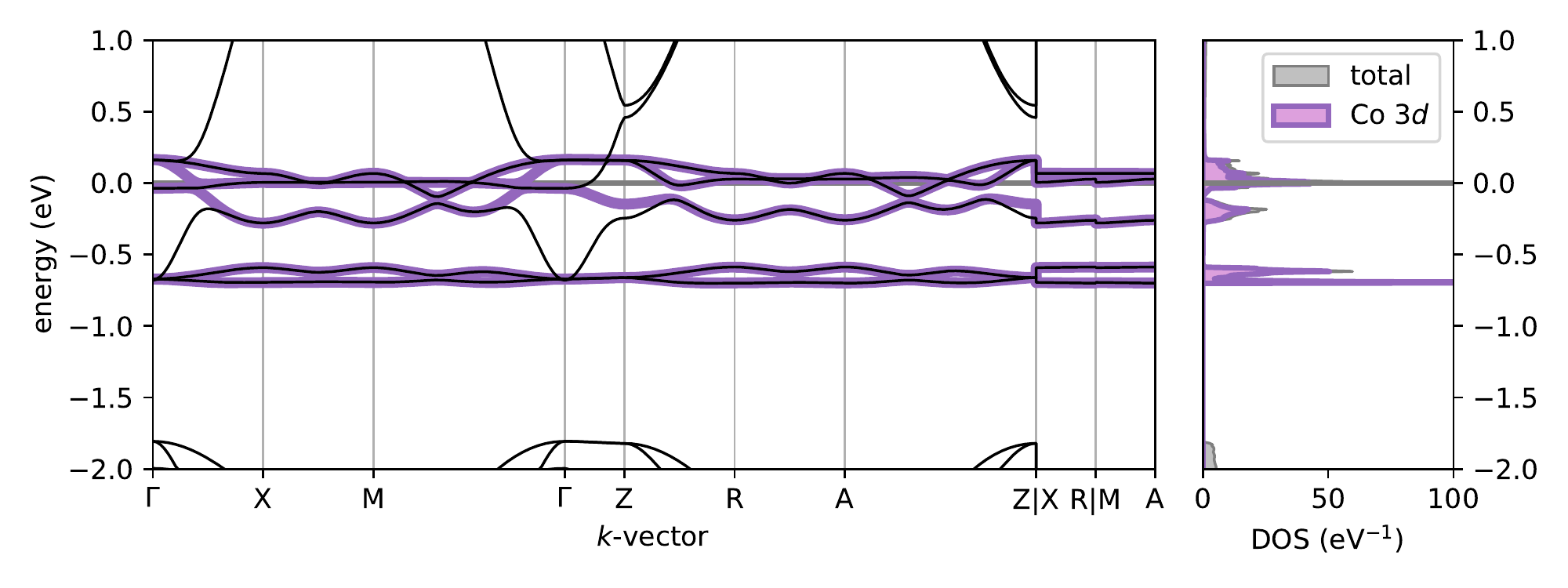}
\caption{\label{zno222r}
Band structure (thin black lines) and Wannier projection (thick purple lines), as well as
densities of states 
for a $\mathrm{CoZn_{15}O_{16}}$ ($2 \times 2 \times 2$)
supercell with relaxed positions of Co and its oxygen neighbors.
}
\end{figure}

ZnO has wurtzite structure (space group $P6/mmm$ \# 186) with lattice
parameters $a=b=3.2427$, $c=5.1948$. A unit cell contains 2 
%{\color{red} \sout{
formula units. 
%} inequivalent ZnO and O sites}.
The Co impurity is modeled here by the supercell method: we perform DFT calculations
for periodic system with unit cells $\mathrm{CoZn_{7}O_{8}}$
(containing 4 $= 2 \times 2 \times 1$ primitive wurtzite unit cells) and
$\mathrm{CoZn_{15}O_{16}}$ (8$= 2 \times 2 \times 2$);  
one Zn atom in the supercell being substituted by Co.
The positions of Co and of the surrounding oxygens were relaxed. 
The  $3d$-states of Co form a narrow band within the energy window 
$-0.75 < E = \varepsilon _F < 0.2$ (see Figure \ref{zno222r}).  

 %%%%%%%%%%%%%%%%%%%%%%%%%%%%%%%%%%%%%%%%%%%%%%%%%%%%%%%%%%%%%%%%%%%%%

For the trigonal field the Hamiltonian in terms of Stevens
operators may be written as
\begin{align}
\hat H_{\mathrm{LF}} &=\hat H_{\mathrm{cub}}+\hat H_{\mathrm{trig}}, \label{d1Stev}\\
\hat{H}_{\mathrm{cub}} &=-\frac{2}{ 3}B^{0}_{4}(\hat{O}_{4}^{0}-20\sqrt{2}
\hat{O}_{4}^{3}) \nonumber \\
\hat{H}_{\mathrm{trig}} &= B^{^{\prime}}_{2} \hat{O}
_{2}^{0}+B_{4}^{^{\prime}}\hat{O}_{4}^{0}
\end{align}
The basis of real functions
%
%\begin{equation}\label{basf}\begin{array}{ll}
\begin{align}
|x\rangle &=\sqrt{\frac{2}{3}}|x^{2}-y^{2}\rangle-\sqrt{\frac{1}{3}}|zx\rangle ,\nonumber\\
|y\rangle &=-\sqrt{\frac{2}{3}}|xy\rangle-\sqrt{\frac{1}{3}}|zy\rangle ,\nonumber \\
|z\rangle &=|z^{2}\rangle ,\label{basf} \\
|v\rangle &=\sqrt{\frac{1}{3}}|x^{2}-y^{2}\rangle+\sqrt{\frac{2}{3}}|zx\rangle ,\nonumber\\
|w\rangle &=-\sqrt{\frac{1}{3}}|xy\rangle+\sqrt{\frac{2}{3}}|zy\rangle \nonumber, 
\end{align}
%\end{array}\end{equation}
diagonalizes the cubic part of $\hat H_{\mathrm{LF}}$ ($C_3$-axis along $z$) with an energy splitting $\Delta$. 
The three basis functions $|x\rangle$,
$|y\rangle$ and $|z\rangle$ build up the $t_{2g}$ representation of
the tetrahedral cubic group and $|v\rangle$, $|w\rangle$ span up the $e_g$
subspace. 
The corresponding diagonal on-site matrix-elements are
$E_z-E_d=(2/5)\Delta -(2/3)v$, $E_x-E_d=E_y-E_d=(2/5)\Delta +v/3$, 
and $E_v-E_d=E_w-E_d=-(3/5) \Delta$, all 
counted from $E_d=(2E_x+E_z+2E_v)/5$, which is the $d$-level energy in the absence of 
ligand field.
The trigonal part $\hat{H}_{\mathrm{trig}}$ splits  the $t_{2g}$ 
level ($E_x-E_z \equiv v$) and has non-diagonal matrix elements 
$\langle x|\hat{H}_{\mathrm{trig}}|v\rangle =
\langle y|\hat{H}_{\mathrm{trig}}|w\rangle \equiv -v^{\prime}$.
\begin{table}[tbp]
\caption{\label{tab:zno} 
Values  (in meV) of the on-site Wannier matrix and the LF parameters
to be used in $\hat H_{\mathrm{LF}}$. The LF parameters obtained by a
Wannier fit are compared with two LF parameter sets obtained by fitting to 
optical spectra \cite{Koidl77} or using a point charge model \cite{Macfarlane70}.
}
\begin{ruledtabular}  
\begin{tabular}{lrrrr}
    &2$\times $2$\times $1, & 2$\times $2$\times $2,& Koidl$^a$  & Macfarlane$^b$ \\
\hline
$E_d-E_F$           &-250.3&-278.6 & & \\
$E_x-E_d$           & 199.5&229.2  & 193.4& 181.8\\ 
$E_z-E_d$           & 281.8&273.7  & 208.3& 231.4\\
$E_v-E_d$           &-340.4&-366.0 &-297.6&-297.6\\
$v^{\prime}$        &-81.1 & -52.7 & -39.7& -43.4\\
$v$                 &-82.3 & -44.5 & -14.9& -49.6\\
$\Delta $           &567.4 & 610.0 & 495.9& 495.9\\
$B^{0}_{4}$         &-4.0  &-4.65  & -3.86& -3.74\\
$B_{4}^{^{\prime}}$ & 1.8  & 1.12  &  0.71& 1.01\\
$B^{^{\prime}}_{2}$ & 7.0  & 4.98  &  4.64& 3.48\\
\hline
$^a$Ref. \onlinecite{Koidl77}\\
$^b$Ref. \onlinecite{Macfarlane70}\\
\end{tabular}
\end{ruledtabular}
\end{table}
%K: Ex,Ez,Ev,vp,v,dlt=193.41640319881 208.294588060257 -297.563697228938 -39.6751596305251 -14.8781848614469 495.939495381563
%B_4^0= -3.85771271673621 B_4^{prime}= 0.707442200854686 B_2^{prime}= 4.63524346743644 meV
%
%Mac: Ex,Ez,Ev,vp,v,dlt=181.844481639907 231.438431178063 -297.563697228938 -43.3947058458868 -49.5939495381563 495.939495381563
%B_4^0= -3.73936240081352 B_4^{prime}= 1.01177158465588 B_2^{prime}= 3.48308635936819 meV

We have constructed Wannier functions via projection on 
the combinations of the $3d$-orbitals given by Eqs.\ (\ref{basf}).
The values of the on-site Wannier matrix are given in the first 
four rows of Table \ref{tab:zno}. The LF parameters are
then found by using the relations (see also \cite{Kuzian06}):
\begin{align}
B^{0}_{4} & = \frac{1}{120}\left(E_v-E_x\right) \approx -4.6 \text{~meV} \nonumber \\
B_{4}^{\prime} & = -\frac{1}{140}\left(E_x-E_z+\frac{3v^{\prime}\sqrt{2}}{2}\right)
\approx 1.1 \text{~meV} \label{E2B} \\
B^{\prime}_{2} & = \frac{1}{21}\left(E_x-E_z-2v^{\prime}\sqrt{2}\right) \approx 5.0  \text{~meV} \nonumber
\end{align}

The Wannier functions of the fully relativistic non spin-polarized GGA functional lead to identical LF
parameters and allow to calculate also the spin-orbit couplings. 
The corresponding on-site matrix shows medium
%The corresponding on-site matrix is given in Appendix 
%\ref{App:SOC} and one observes medium 
deviations from spherical symmetry, in between the weak deviations for 
La$_2$NiO$_4$ and the strong ones for Li$_2$CuO$_2$. Since in the given case of a $d^7$ configuration all
SOC matrix elements are important for the $g$-factors and the multiplet spectrum, we take an average
over the four different numerical values of SOC matrix elements to obtain $\zeta=58.6$ meV, quite close to the 
value of 53.3 meV estimated by Koidl from a fit to the optical absorption spectra \cite{Koidl77}. 

With these LF and SOC parameters, we calculated the multiplet spectrum, the zero field splitting (ZFS), the $g$-factors,
and the optical spectrum with the help of the ELISA code. The less critical Coulomb 
parameters $F^{(2)}=7.65$ eV and 
$F^{(4)}=5.46$ eV were taken as in \cite{Koidl77}. 
These parameters lead to a reasonable agreement with the experimental
multiplet spectrum, as visible in several absorption bands of ZnO:Co \cite{Koidl77}. In Table \ref{TabZnOCo1} we 
compare the average position of each band, i.e.\ the spectrum where SO coupling and trigonal distortion are
neglected. The ELISA code reproduces also the fine structure very well which is demonstrated in Table 
\ref{TabZnOCo2} for the ${}^4 T_2$ absorption band lying in the infrared region.
\begin{table}[h!]
\begin{center}
\caption{
Multiplet energies of Co$^{2+}$ in ZnO as measured by optical absorption \cite{Koidl77} in comparison
to those calculated by the ELISA program with LF parameters from a Wannier fit. Tabulated are the main levels 
without fine structure.}
%, i.e.\ neglecting SO coupling and the trigonal distortion. The energy positions are obtained 
%by taking the mean value of each band.}
%Comparison of multiplet energies fas obtained from the ELISA program with LF parameters from a Wannier fit 
%with quantum chemical calculations (QC) \cite{Xu2019} and experimental RIXS data \cite{Fabbris2017}.  
%quantum chemistry (Xu thesis [2]) and with Wannier LF parameters. The last column compares with the experimental spectrum (see supplemental
%material in [1])
%(Spin-orbit coupling neglected for simplicity).} 
\label{TabZnOCo1}
\vspace*{0.3 cm}
\begin{tabular}{| c | c | c  |} 
\hline
\hline
%$\mathbf{U_{eff}}$ \textbf{(eV)}   & $\mathbf{E_g$} \textbf{spin up (eV)} &  $ \mathbf{E_g}$ \textbf{spin down (eV)} \\
Notation   & Experiment \cite{Koidl77} & This work \\  
   & (eV) & (eV)  \\
\hline
${}^4 A_2$ & 0.00 & 0.00  \\
${}^4 T_2$   &  0.51 & 0.61 \\
${}^4 T_1$  &  0.84 & 1.05  \\
${}^2 E$     &  1.88 &  1.95 \\
${}^4 T_1$   &  2.04 & 2.21 \\
\hline
\hline
\end{tabular}
\end{center}
\end{table}
\begin{table}[h!]
\begin{center}
\caption{
Fine structure of the ${}^4 T_2$ band; comparison between the ELISA multiplet calculation using parameters 
explained in the text and experimental date. The lowest level of the band has been set to zero in both cases.}
%material in [1])
%(Spin-orbit coupling neglected for simplicity).} 
\label{TabZnOCo2}
\vspace*{0.3 cm}
\begin{tabular}{| c | c | c  |} 
\hline
\hline
%$\mathbf{U_{eff}}$ \textbf{(eV)}   & $\mathbf{E_g$} \textbf{spin up (eV)} &  $ \mathbf{E_g}$ \textbf{spin down (eV)} \\
Notation   & This work & Experiment \cite{Koidl77} \\  
   & (meV) & (meV)  \\
\hline
$E_{3/2}$ & 0.0 & 0.00  \\
$E_{1/2}$   &  5.9 & 2.4 \\
$E_{1/2}$  &  17.3 & 12.2  \\
$E_{3/2}$     &  40.5 &  38.2 \\
$E_{1/2}$   &  50.8 & 42.1 \\
$E_{1/2}$   &  63.7 & 55.1 \\
\hline
\hline
\end{tabular}
\end{center}
\end{table}

The lowest ${}^4 A_2$ multiplet of ZnO:Co corresponds to an effective spin 3/2 system with an easy plane anisotropy
\cite{Kuzian06} visible in a zero field splitting (ZFS) between a lower and an upper doublet. In Table 
\ref{ZFS} we compare ZFS and $g$-factors obtained from the ELISA code using Wannier parameters with experimental
results, finding a good agreement in general.

\begin{table}[h!]
\begin{center}
\caption{
Comparison of $g$-factors and ZFS.}
\label{ZFS}
\vspace*{0.3 cm}
\begin{tabular}{| c | c | c |  c |} 
\hline
\hline
%$\mathbf{U_{eff}}$ \textbf{(eV)}   & $\mathbf{E_g$} \textbf{spin up (eV)} &  $ \mathbf{E_g}$ \textbf{spin down (eV)} \\
   & This work & Sati\cite{Sati06} & Koidl\cite{Koidl77} \\  
%   & (meV) & (meV)  \\
\hline
ZFS/cm$^{-1}$ & 5.98 & 5.52 & 5.4 \\
$g_z$ & 2.22 & 2.236 & \\
$g_x=g_y$ & 2.25 & 2.277 & \\
%$E_{3/2}$ & 0.0 & 0.00  \\
%$E_{1/2}$   &  5.3 & 2.4 \\
%$E_{1/2}$  &  15.3 & 12.2  \\
%$E_{3/2}$     &  35.9 &  38.2 \\
%$E_{1/2}$   &  45.6 & 42.1 \\
%$E_{1/2}$   &  57.4 & 55.1 \\
\hline
\hline
\end{tabular}
\end{center}
\end{table}

\begin{figure} [h!]
%%%%%%%%%%%%%%%%%%%%%%%%%%%%%%%%%%%%%%%%%%%%%%%%%%%%%%%%%%%%%%%
%\includegraphics[%draft, 
%width=0.45\textwidth]{fig6a}
%\includegraphics[%draft, 
%width=0.45\textwidth]{fig6b}
\includegraphics[%draft, 
width=0.50\textwidth]{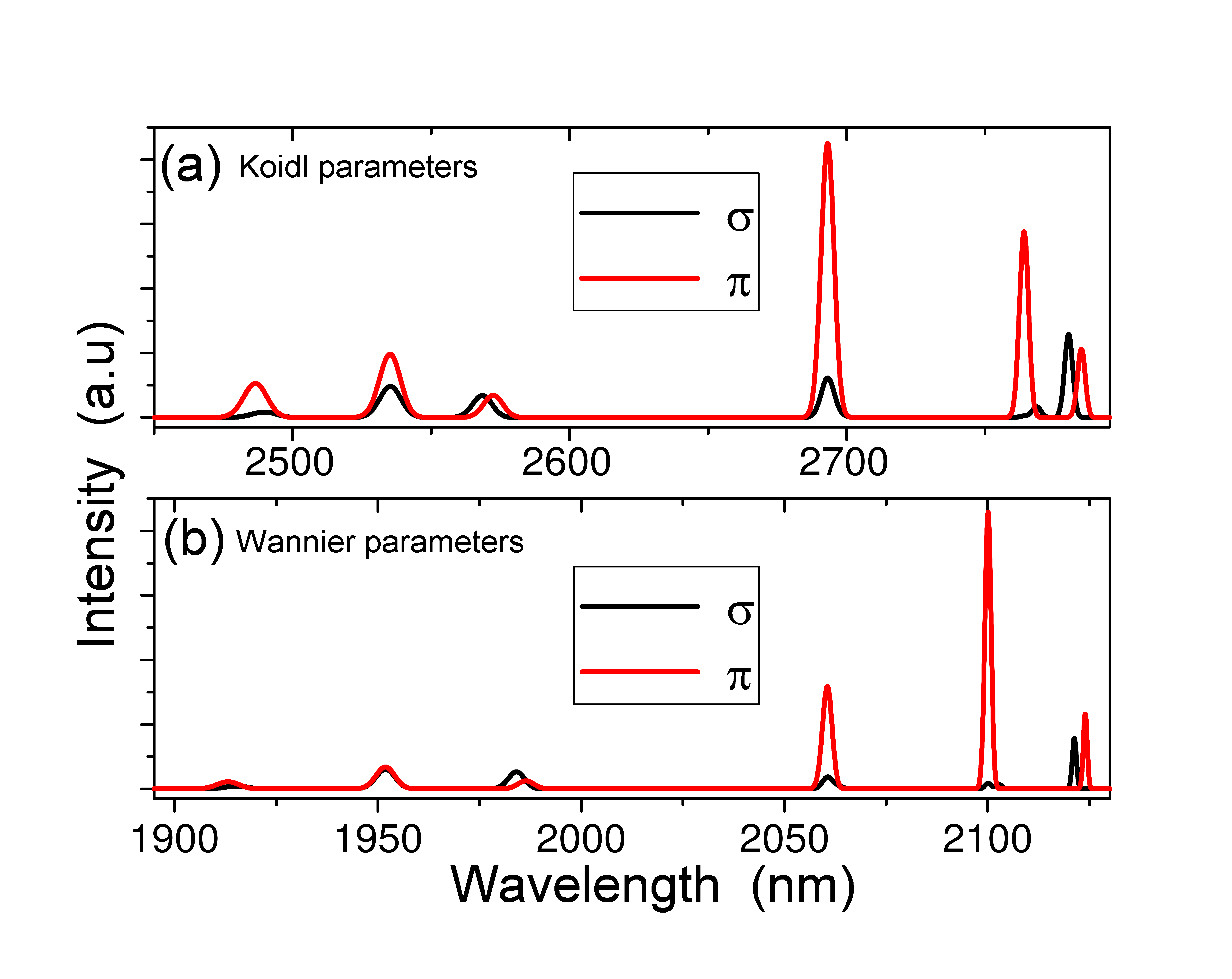}
\caption{\label{opt}
Optical absorption spectra of the ${}^4 T_2$ band with $\pi$- and $\sigma$-polarization as calculated from the 
ELISA code with parameters fitted to the experimental result \cite{Koidl77} (upper panel a) or Wannier parameters
(lower panel b). The width of the Gaussian broadening increases linearly from 0.1 to 1 meV going from
lower to higher energies.}
%The dependence of on-site energies of Wannier functions having $d$-function
%symmetry on the energy window for the 
%high-$T_c$ parent 
%compounds $\mathrm{CaCuO_2}$ (a), $\mathrm{Li_2CuO_2}$ (b) and $\mathrm{ZnO:Co}$ (c). 
%The energy window is defined 
%by a function being 1 between $E_{min}$ and $E_{max}$ and falling of as a Gaussian with a width $dE=1$~eV outside %this window for (a) and (b) and with a width $dE=0.5$~eV for (c).}
\end{figure} %%%%%%%%%%%%%%%%%%%%%%%%%%%%%%%%%%%%%%%%%%%%%%%%%%%%%%%%%%%%%%%%%%%%%

A more precise comparison of the theoretical multiplet spectrum with the optical absorption spectra demands
also the calculation of the dipole matrix elements which is done here following the approach of Sugano and Tanabe
\cite{Sugano70} (\ref{o1}). 
In the case of ZnO:Co the parity breaking perturbation is due to the lack of inversion symmetry
in the tetrahedron and the trigonal distortion. 
That 
perturbation can be expressed in terms of spherical harmonics in the form
\begin{equation}
\label{pbt}
%V_\st{odd} = B_3 (\sqrt{\frac{5}{2}} \hat{O}^0_3 + \hat{O}_3^3).
V_\st{odd} = B_3 \left( \sqrt{\frac{5}{2}} Y_3^0 + Y_3^{+3} + Y_3^{-3} \right) .
\end{equation}
As in the experiment \cite{Koidl77}, we distinguished two polarizations, the linear $\pi$-polarization along
the $z$-axis and the circular polarizations $\sigma_+$ and $\sigma_-$ within the $x$-$y$ plane which are,
however, identical without external magnetic field. We explain in Appendix \ref{Tanabe} how the
dipole matrix element (\ref{o1}) simplifies for these two polarizations and show in Figure \ref{opt} the optical
absorption with $\pi$- and $\sigma$-polarization for the ${}^4 T_2$ band. In the upper panel we choose LF 
parameters 
%{\color{red} (Don't you have taken the LF parameter given by Koidl?)} 
such that they represent the experimental spectra \cite{Koidl77} in an optimal way
($\Delta=465.7$ meV, $v=-14.9$ meV, and $v^{\prime}=-39.7$ meV, 
being slightly improved with respect to the values given in \cite{Koidl77}). 
The resulting curves coincide with the experimental
spectra with respect to the peak positions and the relative heights but cannot reproduce the phonon side-band
which is visible in the experimental spectra between 2.2 and 2.4 $\mu$m wavelength. In the lower panel 
of Fig.\ \ref{opt} we show
the same ${}^4 T_2$ band, but calculated with the Wannier parameter set. We find a good agreement with the 
upper  panel
and with experiment, besides a general shift of about 0.75 $\mu$m due to an overestimation of
the cubic LF splitting $\Delta$ in the Wannier fit.

%multiplet spectrum, presented first without fine structure (Table \ref{TabZnOCo1}), and then the fine structure of
%the ${}^4 T_2$ band in Table \ref{TabZnOCo2}. 

\subsection{Actinide oxyde UO$_2$}
\begin{table}[tbp]
\caption{\label{tab:UO2} 
The parameters and resulting energies (in meV) for UO$_2$.  
}
\begin{ruledtabular}  
\begin{tabular}{crrrr}
 & This work & LDA+$U^a$ & INS$^b$ & INS$^c$  \\
\hline
$V_4$           & -146  & -93 &  -123& -116.2\\
$V_6$           & 34    & 16  &  26.5& 25.8\\
$F^2$           & 5339.3& 5649& 5339.3& \\
$F^4$           & 4562.9& 3774& 4562.9& \\
$F^6$           & 3607.2& 2791& 3607.2& \\
$\zeta$         &  222.7&  230&  222.7& \\
$\Gamma _5$     &     0&     0&     0& 0 \\
$\Gamma _3$    & 165.1& 125.7& 150.1& 150\\
$\Gamma _4$    & 169.7& 156.3& 166.7& 158\\
$\Gamma _1$    & 175.5& 174.0& 174.8& 170\\
\hline
$^a$Ref. \onlinecite{Zhou2012}\\
$^b$Ref. \onlinecite{Amoretti1989}\\
$^c$Ref. \onlinecite{Nakotte2010}
\end{tabular}
\end{ruledtabular}
\end{table}

To test the applicability of our approach to 
a $5f$-electron system, we have chosen 
UO$_2$, whose importance as a nuclear fuel makes it
an object of intensive studies.
It is an antiferromagnetic insulator. But that is not correctly reproduced
by standard spin-polarized DFT calculations without Hubbard $U$ correction (LSDA or 
SGGA functionals) \cite{Crocombette2001} who find metallic behavior. To obtain correct total energies
for the study of point defects, diffusion, or thermodynamic properties it is important
to combine the DFT+$U$ functionals with the occupation matrix control \cite{Dorado2009} to 
avoid spurious local minima. In such a way, also the LF parameters of UO$_2$ had 
been calculated modifying the standard SGGA+$U$ functional \cite{Zhou2012}. But we will show now
that the much more simple Wannier function method leads already to correct results.

The U$^{4+}$ ion has $5f^2$ configuration and is surrounded by 
four oxygen ions forming an ideal tetrahedron. The ligand field has a cubic symmetry
and is described by the effective Hamiltonian
\begin{align}
\hat H_{\mathrm{LF,c}} &= B_{4}\left(\hat{O}_{4}^{0}+5\hat{O}_{4}^{4}\right)
 + B_{6}\left(\hat{O}_{6}^{0}-21\hat{O}_{6}^{4}\right). \label{Hcubf}
\end{align}
The common notations for its two parameters are \cite{Abragam,Amoretti1989}
\begin{align}
B_{4} &\equiv V_4\beta ,\quad \beta = \frac{2}{11\cdot 45} , \\
B_6 &\equiv V_6\gamma ,\quad \gamma = -\frac{4}{11\cdot 13\cdot 27} .
\end{align}
%beta = 2/(11*45.0); gamma = -4/(11*13*27.0);
To perform the Wannier fit we use the non spin-polarized GGA bandstructure of UO$_2$ (not shown).
There, the 5$f$ bands are localized close to the Fermi energy and well separated from the rest
of the spectrum. Using an energy window between -1.0 and 2.2 eV we find the   
 Wannier function on-site matrix given in Table \ref{tab:HU}.
The Wannier functions are found by projecting onto $5f$ real spherical harmonics, as
defined in Ref. \onlinecite{Koepernik99}. 
The on-site Wannier matrix can be easily diagonalized leading to the exact analytical eigenvalues and 
eigenfunctions given in \cite{Abragam}. The eigenenergies of the on-site Wannier matrix in the 
notation of \cite{Abragam} are (in meV)
%The
%eigenenergies of this Hamiltonian are 
%E0=1.212025757413731;EV=0.318303527772461;EVp=0.393946556035288;
$E(\Gamma _2)=1212.0+\varepsilon_F$, 
$E(\Gamma _5)\equiv E(\Gamma _2)+V=318.3+\varepsilon_F$ and 
$E(\Gamma _4)\equiv E(\Gamma _2)+V^{\prime}=393.9+\varepsilon_F$. 
From these energy differences one can calculate the LF parameter 
values  using the exact expressions of \cite{Abragam}:
%V=EV-E0;Vp=EVp-E0;
%B40=(-V+3*Vp)/(60*44.0); B44=5*B40;
%B60=(9*V-5*Vp)/(180*616.0); B64=-21*B60;
\begin{align}
B_{4} &= \frac{-V+3V^{\prime}}{2640}, \\
B_6 &= \frac{9V-5V^{\prime}}{110880}.
\end{align}
The so obtained LF parameters are compared in Table \ref{tab:UO2} with those obtained 
by the elaborated LSDA+$U$ calculation with occupation matrix control \cite{Zhou2012}, and
with parameter sets due to fitting to inelastic neutron scattering experiments 
\cite{Amoretti1989,Nakotte2010}. The Wannier LF parameter are
then also used in the ELISA program to calculate the multiplet structure of the 5$f^2$ configuration.
For simplicity we used the same SO and Coulomb parameter as in \cite{Amoretti1989}. The lowest
multiplet of the 5$f^2$ configuration is ${}^3H_4$ which is split by the cubic ligand field into 
the levels given in Table \ref{tab:UO2}. We find a very good agreement despite the numerical simplicity of our
approach.

%Table \ref{tab:UO2} compares the parameters and energy levels obtained 
%by elaborated LDA+$U$ calculations \cite{Zhou2012} and in inelastic 
%neutron scattering experiments \cite{Amoretti1989,Nakotte2010}. Our 
%simple approach 
%gives surprisingly good agreement with these data.

\section{Discussion}
%{\color{red} (RH: Very preliminary version.)} 
The good agreement between our {\em{ab-initio}} calculated LF parameters and experimentally well established values 
for the treated examples demonstrates that there are indeed only two main contributions to the LFs: the 
electrostatic potential in the crystal and hybridization to the neighboring ligands 
(see also \cite{Savoyant2009,Savoyant2010}).
%{\color{red}(We can cite here papers of Anatoli and I, and Roman and I  dealing with hybridization and electrostatic contribution to the LF parameter: savoyant et al, PRB 80, 115203 (2009), and  savoyant et al, phys. status solidi 247, 1961 (2010))}. 
One can also conclude that the 
electron-electron interaction has only a small influence. There is only one exception in the list of materials which we treated:
it is ZnO:Co where the calculated cubic splitting $\Delta$ (or $B_4^0$) exceeds the experimental value considerably. 
The cubic splitting is mainly determined by the energy distance between $d$ and $p$ levels \cite{Kuzian06} 
which is probably not correctly calculated  in GGA. It was shown by the authors of Ref.\  \cite{Novak13} that the relative position of the ligand
levels with respect to the $d$ or $f$ bands is influenced by electron correlation effects for which they used this 
energy distance as a free parameter in their method to determine CF (or LF) parameters. 

Due to the rather simple origin of the the LF parameters it can also be expected that the filling of the $d$ or $f$ 
shell in otherwise similar compounds does not drastically change the  LF parameters, reflecting their one-electron character. In that respect it is instructive
to compare the two treated tetragonal cases, La$_2$NiO$_4$ and CaCuO$_2$ which show a similar $d$-level 
ordering with one exception: the position of the $d_{z^2}$ and $d_{xy}$ orbitals are exchanged. That can be explained 
by the absence of the apex oxygen in the infinite layer compound CaCuO$_2$. Since it is just this apex oxygen 
which pushes the $d_{z^2}$ level higher in energy such that it becomes most close to the $d_{x^2-y^2}$ orbital.

The results of our method depend in a sensitive way on the choice of the energy window for the Wannier fit. The dependence is less critical
when the $d$ or $f$ band is well separated from the rest of the band structure as for UO$_2$, or when the overlap is small as for La$_2$NiO$_4$
or ZnO:Co. In these situations the energy window has just to encompass the relevant bands and increasing its width does not alter 
the LF parameters in a sensitive case as it is shown in Appendix \ref{App:chewin} for ZnO:Co. The choice of the energy window is more
critical when there is no clear separation as for CaCuO$_2$. In that case we fix the lower limit of the energy window such that it separates
the antibonding $d$-like states from the bonding ones (see Appendix \ref{App:chewin}), but there remains 
an error of the Wannier matrix elements $H_{ii}$ which we estimate
to be about  0.1 eV. In the case of a good separation the error is about ten times smaller.

%as long as the energy window does no
%Precision of LF determination sensibility to chosen energy window. Less critical when $d$ band is well separated, 
%very critical if there is no separation as for CaCuO$_3$. Estimated error $E_i-E_j \propto 0.1$ eV.

We call our method \emph{ab initio} but we do not understand that in the most strict sense. So, we propose to add a level broadening by hand to simplify 
the comparison with the experimental spectra. Also, we have shown that, occasionally, an adjustable exchange field improves 
the RIXS spectra for cuprates. This latter should be understood as arising 
from the antiferromagnetic nearest neighbor exchange couplings in the CuO$_2$ plane which are not explicitly included in our 
Hamiltonian, but nevertheless 
physically very well justified.

%RIXS: exchange field improves the results for cuprates. 

A possible improvement of our method is to relax the condition of spherical symmetry for SOC first, and after that also 
for the Coulomb interaction, two 
secondary effects of reducing the symmetry in the crystal.
That introduces many new parameters, but the example of Li$_2$CuO$_2$ 
shows that they can be calculated from the Wannier fit. To perform an analogous procedure for the Coulomb interaction
demands more work.

The method which we propose is close in spirit to \cite{Novak13} but applied here to less correlated 3$d$ and 5$f$ electrons in
contrast to the 4$f$ systems there. Treating the energy difference to the ligand levels as free parameter might sometimes improve
the resulting LF parameters. However, in most cases which we considered, we found such a correction not necessary, besides for ZnO:Co as mentioned above. 
Importantly, we do not only calculate the LF parameter, but also the multiplet spectra and many spectroscopical curves. That is similar to \cite{Haverkort2012} where, 
however, the ligand orbitals are explicitely treated in the exact diagonalization procedure to calculate the multiplet spectra which breaks the relation to the 
traditionally known LF parameters.

\section{Conclusion}
We present a simple, general, and precise method of calculations of the multiplet spectrum and the 
relevant experimental spectra of local magnetic $d$- and $f$-centers in solids in an \emph{ab initio} manner. 
The method combines the 
Wannier analysis of the nonmagnetic GGA band-structure with an exact diagonalization method
of the local, atomic like, multiplet Hamiltonian containing Coulomb, spin orbit, and ligand field interactions.  
%{\color{red} (RH: to be continued: experiments ...., we tested .... .)} 
Despite its simplicity our method is remarkably precise. That gives us 
confidence to predict the multiplet structure of less
well known systems. The precision we obtained is sufficient to classify the relevant spectroscopic terms.

\begin{acknowledgments}
This work was supported by the National Academy of Sciences
of Ukraine (Project No. III-4-19).
We especially thank Manuel Richter and Liviu Hozoi for detailed discussions on the subject. We also thank Shi Lei, Ibrahim Mansouri, and Maen Salman for preliminary calculations at an early stage of the project, as well as Ulrike Nitzsche, Claude Arnold, and 
Andrey Likhtin for technical assistance, and the IFW Dresden (Germany) for the use of their computer facilities. 
We had valuable discussions with Valentina Bisogni, Yuri Ralchenko, 
Michael Kuzmin, and Michel Freyss on several subjects of the presented work, for which we are very grateful. O.J. was supported by the Leibniz Association
through the Leibniz Competition. 
\end{acknowledgments}

\newpage

\appendix
\begin{widetext}

\section{\emph{Ab initio} determination of spin-orbit coupling parameters}
\label{App:Wsoc}

\subsection{On-site Hamiltonians $H_{ii}$ obtained by wannierization of full-relativistic band-structures of \lanio}\label{App:Hii}

\begingroup
\squeezetable
\begin{table}[h!]
%\begin{table}[htbp]
\caption{\label{tab:Hii_lanio_exp}
On-site Hamiltonian $H_{ii}$ obtained by wannierization of the full-relativistic
band-structure for the experimental structure of \lanio.  All values are in meV.
}
\begin{ruledtabular}
\begin{tabular}{r|ccccc|ccccc}
        & \ketxyu & \ketyzu & \ketzru & \ketxzu & \ketxxu & \ketxyd & \ketyzd & \ketzrd & \ketxzd & \ketxxd \\ \hline
\braxyu & $-1250.1$ & 21.9 & $-0.2i$ & $-4.3i$ &$69.7i$& 0.0 & $-26.0\!-\!26.0i$ & 0.0 & $-25.7\!+\!25.7i$ & $-5.4\!+\!5.4i$ \\
\brayzu & 21.9 & $-1193.7$ & $-6.0i$ &$39.0i$& $-2.9i$ &$26.0\!+\!26.0i$& 0.0 & $-45.5\!+\!45.5i$ & $-2.4\!+\!2.4i$ & $-25.3\!+\!25.3i$ \\
\brazru &$0.2i$&$6.0i$& $-130.6$ & 3.0 & 6.1 & 0.0 & 45.5$-45.5i$ & 0.0 &$45.9\!+\!45.9i$& 0.0 \\
\braxzu &$4.3i$& $-39.0i$ & 3.0 & $-1195.8$ & $-0.5$ & 25.7$-25.7i$ & 2.4$-2.4i$ & $-45.9\!-\!45.9i$ & 0.0 &$25.6\!+\!25.6i$\\
\braxxu & $-69.7i$ & 2.9i & 6.1 & $-0.5$ & 333.4 & 5.4$-5.4i$ & 25.3$-25.3i$ & 0.0 & $-25.6\!-\!25.6i$ & 0.0 \\ \hline
\braxyd & 0.0 & 26.0$-26.0i$ & 0.0 &$25.7\!+\!25.7i$&$5.4\!+\!5.4i$& $-1250.1$ & 21.9 &$0.2i$&$4.3i$& $-69.7i$ \\
\brayzd &$-26.0\!+\!26.0i$ & 0.0 &$45.5\!+\!45.5i$&$2.4\!+\!2.4i$& $25.3\!+\!25.3i$ & 21.9 & $-1193.7$ &$6.0i$& $-39.0i$ &$2.9i$\\
\brazrd & 0.0 & $-45.5\!-\!45.5i$ & 0.0 & $-45.9\!+\!45.9i$ & 0.0 & $-0.2i$ & $-6.0i$ & $-130.6$ & 3.0 & 6.1 \\
\braxzd &$-25.7\!-\!25.7i$ & $-2.4\!-\!2.4i$ & 45.9$-45.9i$ & 0.0 & $-25.6\!+\!25.6i$ & $-4.3i$ &$39.0i$& 3.0 & $-1195.8$ & $-0.5$ \\
\braxxd & $-5.4\!-\!5.4i$ & $-25.3\!-\!25.3i$ & 0.0 & 25.6$-25.6i$ & 0.0 &$69.7i$& $-2.9i$ & 6.1 & $-0.5$ & 333.4 \\
\end{tabular}
\end{ruledtabular}
\end{table}
\endgroup
%\end{widetext}

%\begingroup
%\squeezetable
\begin{table}[h!]
%\begin{table}[htbp]
\caption{\label{tab:Hii_lanio_simpl}
On-site Hamiltonian $H_{ii}$ obtained by wannierization of the full-relativistic
band-structure for the simplified structure of \lanio.  All values are in meV.
}
\begin{ruledtabular}
\begin{tabular}{r|ccccc|ccccc}
        & \ketxyu & \ketyzu & \ketzru & \ketxzu & \ketxxu & \ketxyd & \ketyzd & \ketzrd & \ketxzd & \ketxxd \\ \hline
\braxyu &$-1254.9$ & 0.0 & 0.0 & 0.0 & $69.2i$ & 0.0 & 36.8 & 0.0 & $-36.8i$ & 0.0 \\
\brayzu &0.0 & $-1200.4$ & 0.0 & $38.9i$ & 0.0 & $-36.8$ & 0.0 & $-64.3i$ & 0.0 & $-35.3i$ \\
\brazru &0.0 & 0.0 & $-131.2$ & 0.0 & 0.0 & 0.0 & $64.3i$ & 0.0 & $-64.3$ & 0.0 \\
\braxzu &0.0 & $-38.9i$ & 0.0 & $-1200.4$ & 0.0 & $36.8i$ & 0.0 & 64.3 & 0.0 & $-35.3$ \\
\braxxu &$-69.2i$ & 0.0 & 0.0 & 0.0 & 329.9 & 0.0 & $35.3i$ & 0.0 & 35.3 & 0.0 \\ \hline
\braxyd &0.0 & $-36.8$ & 0.0 & $-36.8i$ & 0.0 & $-1254.9$ & 0.0 & 0.0 & 0.0 & $-69.2i$ \\
\brayzd &36.8 & 0.0 & $-64.3i$ & 0.0 & $-35.3i$ & 0.0 & $-1200.4$ & 0.0 & $-38.9i$ & 0.0 \\
\brazrd &0.0 & $64.3i$ & 0.0 & 64.3 & 0.0 & 0.0 & 0.0 & $-131.2$ & 0.0 & 0.0 \\
\braxzd &$36.8i$ & 0.0 & $-64.3$ & 0.0 & 35.3 & 0.0 & $38.9i$ & 0.0 & $-1200.4$ & 0.0 \\
\braxxd &0.0 & $35.3i$ & 0.0 & $-35.3$ & 0.0 & $69.2i$ & 0.0 & 0.0 & 0.0 & 329.9 \\
\end{tabular}
\end{ruledtabular}
\end{table}
%\endgroup

\subsection{Spin-orbit interaction}
\label{App:SOC}
The spin-orbit interaction Hamiltonian (\ref{eq:socZ}) may be rewritten as 
 \begin{equation}
\hat{H}_{\mathrm{SOC}} = 
 \frac{\zeta }{2}\sum_i \left[2\hat{l}_i^{z}\hat{s}_i^{z}+
  \hat{l}_i^{+}\hat{s}_i^{-}+\hat{l}_i^{-}\hat{s}_i^{+}\right],\label{eq:soc}
\end{equation}
where the summation goes over electrons. 
We will write its matrix in the basis of real spherical harmonics $\left|m,\sigma \right\rangle$,
where  $m= xy, yz, 3z^2-r^2, xz, x^2-y^2$, $\sigma = \pm \nicefrac{1}{2}$. 
The matrix element values
% \begin{align}
%\left\langle m,\sigma \right|& 2\hat{l}^{z}\hat{s}^{z}+
 %\hat{l}^{+}\hat{s}^{-}+\hat{l}^{-}\hat{s}^{+}\left|m^{\prime},\sigma^{\prime}\right\rangle = %\nonumber \\
 %& 
 %2\sigma \delta _{\sigma,\sigma^{\prime}}\left\langle m,\sigma\right|\hat{l}^{z}\left|m^{\prime},\sigma \right\rangle + \nonumber \\
%&  +\delta _{\sigma,-\nicefrac{1}{2}}\delta _{\sigma^{\prime},\nicefrac{1}{2}}
 %\left\langle m,-\nicefrac{1}{2}\right|\hat{l}^{z}\left|m^{\prime},-\nicefrac{1}{2}\right\rangle %\nonumber \\
 %&
 %+\delta _{\sigma,\nicefrac{1}{2}}\delta _{\sigma^{\prime},-\nicefrac{1}{2}}
 %\left\langle m,\nicefrac{1}{2}\right|\hat{l}^{z}\left|m^{\prime}, \nicefrac{1}{2}\right\rangle
 %\label{eq:socm}
 %\end{align}
%RH: I replaced $m_s$ by $\sigma$ to unify the notation but to my mind there is an error in the above formula which I corrected below. Oleg, please
%check.
\begin{equation}
\left\langle m,\sigma \right| 2\hat{l}^{z}\hat{s}^{z}+
 \hat{l}^{+}\hat{s}^{-}+\hat{l}^{-}\hat{s}^{+}\left|m^{\prime},\sigma^{\prime}\right\rangle =
 2\sigma \delta _{\sigma,\sigma^{\prime}}\left\langle m \right|\hat{l}^{z}\left|m^{\prime} \right\rangle 
  +\delta _{\sigma,-\nicefrac{1}{2}}\delta _{\sigma^{\prime},\nicefrac{1}{2}}
 \left\langle m,\right|\hat{l}^{+}\left|m^{\prime}\right\rangle 
 +\delta _{\sigma,\nicefrac{1}{2}}\delta _{\sigma^{\prime},-\nicefrac{1}{2}}
 \left\langle m\right|\hat{l}^{-}\left|m^{\prime}\right\rangle
 \label{eq:socm}
\end{equation}
%are given in Table \ref{tab:SOC}. 
%to obtain the matrix 
\begin{table*}[h!]
\caption{\label{tab:SOC} The matrix of SO interaction in the basis $\left|m,m_s\right\rangle $ }
\begin{tabular}{r|ccccc|ccccc}
\hline 
 & $\left|xy,\uparrow\right\rangle $ & $\left|yz,\uparrow\right\rangle $ & $\left|3z^2-r^2,\uparrow\right\rangle $ 
 & $\left|xz,\uparrow\right\rangle $  & $\left|x^2-y^2,\uparrow \right\rangle $ 
 & $\left|xy,\downarrow\right\rangle $ & $\left|yz,\downarrow\right\rangle $ & $\left|3z^2-r^2,\downarrow\right\rangle $ 
 & $\left|xz,\downarrow\right\rangle $ & $\left|x^2-y^2,\downarrow\right\rangle $  \\
\hline 
$\left\langle xy,\uparrow\right| $   & 0& 0& 0& 0& $2i$ & 0& 1& 0& $-i$ &0 \\
%\hline 
$\left\langle yz,\uparrow\right| $  & 0& 0& 0& $i$& 0& -1& 0& $-i\sqrt{3}$& 0& $-i$\\
%\hline 
$\left\langle 3z^2-r^2,\uparrow\right| $  & 0& 0& 0& 0& 0& 0& $i\sqrt{3}$& 0& $-\sqrt{3}$& 0\\
%\hline 
$\left\langle xz,\uparrow\right| $  & 0& $-i$& 0& 0& 0& $i$& 0& $\sqrt{3}$& 0& -1\\
%\hline 
$\left\langle x^2-y^2,\uparrow \right| $  & $-2i$& 0& 0& 0& 0& 0& $i$& 0& 1& 0\\
\hline 
$\left\langle xy,\downarrow\right| $  & 0& -1& 0& $-i$& 0& 0& 0& 0& 0& $-2i$\\
%\hline 
$\left\langle yz,\downarrow\right| $  & 1& 0& $-i\sqrt{3}$& 0& $-i$& 0&0 & 0& $-i$& 0\\
%\hline 
$\left\langle 3z^2-r^2,\downarrow\right| $  & 0& $i\sqrt{3}$& 0& $\sqrt{3}$& 0& 0& 0& 0& 0& 0\\
%\hline 
$\left\langle xz,\downarrow\right| $  & $i$& 0& $-\sqrt{3}$& 0& 1& 0& $i$& 0& 0& 0\\
%\hline 
$\left\langle x^2-y^2,\downarrow\right| $  & 0& $i$& 0& -1& 0& $2i$& 0& 0& 0& 0\\
%\hline
\end{tabular}
\end{table*}
\begin{table*}
\caption{\label{tab:Wrel} The matrix of the Hamiltonian (in meV) in the basis of Wannier 
functions obtained in the full-relativistic calculation for $\mathrm{Li_2CuO_2}$. }
\begin{tabular}{r|ccccc|ccccc}
\hline 
 & $\left|xy,\uparrow\right\rangle $ & $\left|yz,\uparrow\right\rangle $ & $\left|3z^2-r^2,\uparrow\right\rangle $ 
 & $\left|xz,\uparrow\right\rangle $  & $\left|x^2-y^2,\uparrow \right\rangle $ 
 & $\left|xy,\downarrow\right\rangle $ & $\left|yz,\downarrow\right\rangle $ & $\left|3z^2-r^2,\downarrow\right\rangle $ 
 & $\left|xz,\downarrow\right\rangle $ & $\left|x^2-y^2,\downarrow\right\rangle $  \\
\hline 
$\left\langle xy,\uparrow\right| $   & $-171.31$& 0& 0& 0& $62.20i$ & 0& $27.14$& 0& $-29.49i$ &0 \\
%\hline 
$\left\langle yz,\uparrow\right| $  & 0& $-2135.16$& 0& $39.44i$& 0& $-27.14$& 0& $-75.896i$& 0& $-43.74i$\\
%\hline 
$\left\langle 3z^2-r^2,\uparrow\right| $  & 0& 0& $-2456.79$& 0& 0& 0& $75.896i$& 0& $-75.83$& 0\\
%\hline 
$\left\langle xz,\uparrow\right| $  & 0& $-39.43i$& 0& $-1989.89$& 0& $29.49i$& 0& $75.83$& 0& $-44.24$\\
%\hline 
$\left\langle x^2-y^2,\uparrow \right| $  & $-62.20i$& 0& 0& 0& $-2075.00$& 0& $43.74i$& 0& $44.24$& 0\\
\hline 
$\left\langle xy,\downarrow\right| $  & 0& $-27.14$& 0& $-29.49i$& 0& $-171.31$& 0& 0& 0& $-62.20i$\\
%\hline 
$\left\langle yz,\downarrow\right| $  & $27.14$& 0& $-75.896i$& 0& $-43.74i$& 0& $-2135.15$& 0& $-39.44i$& 0\\
%\hline 
$\left\langle 3z^2-r^2,\downarrow\right| $  & 0& $75.896i$& 0& $75.83$& 0& 0& 0& $-2456.79$& 0& 0\\
%\hline 
$\left\langle xz,\downarrow\right| $  & $29.49i$& 0& $-75.83$& 0& $44.24$& 0& $39.44i$& 0& $-1989.89$& 0\\
%\hline 
$\left\langle x^2-y^2,\downarrow\right| $  & 0& $43.74i$& 0& $-44.24$& 0& $62.20i$& 0& 0& 0& $-2075.00$\\
%\hline
\end{tabular}
\end{table*}
are given in Table \ref{tab:SOC}.  
This matrix we compare with the non-diagonal part of the Hamiltonian matrix in the basis of Wannier 
functions obtained in the full-relativistic calculation for \lanio (see previos subsection) and $\mathrm{Li_2CuO_2}$ 
(Table \ref{tab:Wrel}). 
%In the effective spin-Hamiltonian one usually 
%
We may see that the full matrix 
of Li$_2$CuO$_2$ 
cannot be described by a single parameter $\zeta$ because the Wannier functions are not spherically 
symmetrical and has contributions from different sites. But to calculate the response of the system to the application of a 
magnetic field we need only the ground state Kramers doublet. This doublet
contains the $d_{xy}$ orbital with an admixture of other $d_m$ orbitals coupled 
to it by the spin-orbit interaction.  The exact diagonalization of the matrix gives the doublet wave
functions
\begin{align}
%[ 0.9993+0.j  -0.+0.j  0.-0.0001j  0.+0.j  -0.-0.0319j   0.-0.j  0.0134-0.j  0.+0.j  -0.+0.0157j  0.+0.j ]
\left|\psi _0,\uparrow\right\rangle &= 0.9993\left|xy,\uparrow\right\rangle 
  -0.0319i\left|x^2-y^2,\uparrow \right\rangle %\nonumber \\
  %&
  +0.0134\left|yz,\downarrow\right\rangle 
  + 0.0157i\left|xz,\downarrow\right\rangle , \\
%[ 0.000e+00+0.j   1.340e-02-0.j   0.000e+00-0.j  -1.000e-04-0.0157j -0.000e+00-0.j   -9.993e-01+0.0033j 0.000e+00+0.j     -0.000e+00-0.0001j -0.000e+00+0.j  -1.000e-04-0.0319j]
\left|\psi _0,\downarrow\right\rangle &= 0.9993\left|xy,\downarrow\right\rangle 
  +0.0319i\left|x^2-y^2,\downarrow\right\rangle %\nonumber \\
    %&
    -0.0134\left|yz,\uparrow\right\rangle 
  +0.0157i\left|xz,\uparrow\right\rangle .
\end{align}
The $g$-factors entering in the
effective spin-Hamiltonian (\ref{eq:Hs})
are then calculated as
\begin{align}
g_{\alpha} &= 2\left|\left\langle \psi _0,\uparrow \right|\hat{l}^{\alpha}
  +g_s\hat{s}^{\alpha}\left|\psi _0,\uparrow \right\rangle\right| .
\end{align}

\subsection{Rotation of the atomic spin-orbit coupling matrix}\label{App:LSRot}
While the rotation of the full 10$\times$10 matrix is cumbersome, it can be
largely simplified if we restrict ourselves to the $t_{2g}$ basis states. Here,
the respective basis transformation reduces to the rotation of orbital angular
momentum matrices $\bf{L'}\xrightarrow{\bf R}\bf{L}$ which can be done using conventional rotation
matrices ${\bf R}$ (here, we rotate by $\frac34\pi$ around the $z$ axis):

%\begin{widetext}
\begin{equation}
H^{t_{2g}}_{\text{SOC}} = \zeta\mathbf{L'}\cdot\mathbf{S} =  
\zeta\left(\mathbf{R}_{\frac{3\pi}{4}z}\mathbf{L}\mathbf{R}^{\sf T}_{\frac{3\pi}{4}z}\right)\cdot\mathbf{S} = 
\frac{\zeta}{2} 
\left(\begin{array}{ccc|ccc}
0  & 0  & 0  & 0  &  -\frac{\sqrt{2}}{2}-\frac{\sqrt{2}}{2}i &  -\frac{\sqrt{2}}{2}+\frac{\sqrt{2}}{2}i \\
0  & 0  &  i  & \frac{\sqrt{2}}{2}+\frac{\sqrt{2}}{2}i & 0  & 0\\
0  &  -i & 0  & \frac{\sqrt{2}}{2}-\frac{\sqrt{2}}{2}i & 0  & 0\\ \hline
0  & \frac{\sqrt{2}}{2}-\frac{\sqrt{2}}{2}i & \frac{\sqrt{2}}{2}+\frac{\sqrt{2}}{2}i & 0 & 0 & 0 \\
 -\frac{\sqrt{2}}{2}+\frac{\sqrt{2}}{2}i & 0  & 0  & 0  & 0 &  -i \\
 -\frac{\sqrt{2}}{2}-\frac{\sqrt{2}}{2}i & 0  & 0  & 0  & i &   0 \\
\end{array}\right).
\end{equation}
%\end{widetext}

Here, ${\bf L}$ and ${\bf S}$ are orbital and spin angular momentum operators
in the matrix form. The rotation matrix is given in the $\left(xy, yz,
xz\right) = \left(z, x, y\right)$ basis and for the $\frac34\pi$ rotation around $z$ equals to:

\begin{equation}
\mathbf{R}_{\frac{3\pi}{4}z}=\left(\begin{array}{ccc}
1 & 0 & 0 \\
0 & -\frac{\sqrt{2}}{2} & \frac{\sqrt{2}}{2} \\
0 & -\frac{\sqrt{2}}{2} & -\frac{\sqrt{2}}{2} \\
\end{array}\right).
\end{equation}

\section{Stevens operators}\label{App:Stev}

%{\color{blue}(RH: I propose to cancel that Appendix. Since the Steven's operators are very well known, they are given in Abragam/Bleaney and they may even be found on wikipedia.)} 
Traditionally, the one-particle ligand field Hamiltonian is expressed 
in terms of Stevens equivalent operators \cite{Abragam}.
Here we give the formulas for operators that was used in this work. 
For $d$-electron systems the ligand field is expressed via the operators 
of second 
\begin{equation}
\hat{O}_{2}^{0}  =   3 \hat{l}_{z}^{2} - l(l+1), \quad 
\hat{O}_{2}^{2} = \frac{1}{2}\left(\hat{l}_{+}^{2} +\hat{l}_{-}^{2} \right), \label{O22} 
\end{equation}
and fourth order 
\begin{align}
\hat{O}_{4}^{2} & = 
 \frac{1}{4}\left\{\left[7\hat{l}_{z}^{2} - l(l+1) -5\right]\left(\hat{l}_{+}^{2} +\hat{l}_{-}^{2} \right) %\right. \nonumber \\
  %& 
  + %\left. 
  \left(\hat{l}_{+}^{2} +\hat{l}_{-}^{2} \right)\left[7\hat{l}_{z}^{2} - l(l+1) -5\right]\right\},  \label{O42} \\
  \hat{O}_{4}^{0} & =   35 \hat{l}_{z}^{4} -30 l(l+1) \hat{l}_{z}^{2} +25 \hat{l}_{z}^{2} %\\
 %&
 -6l(l+1)+3l^2(l+1)^2, \label{O40} \\
  \hat{O}_{4}^{4} & =    \frac {1}{2}\left( \hat{l}_{+}^{4} +\hat{l}_{-}^{4} \right),
\label{O44} \\
\hat{O}_{4}^{3} & =  \frac {1}{4} \left\{ \hat{l}_{z} (\hat{l}_{+}^{3} +\hat{l}_{-}^{3} )
+  \left(\hat{l}_{+}^{3} +\hat{l}_{-}^{3} \right)\hat{l}_{z} \right\}, \label{O43} 
\end{align}
where $\hat{l}_{z}$,$\hat{l}_{+}$, and $\hat{l}_{-}$ are
angular momentum operators.
For $f$-electron systems the sixth order operators appear
\begin{align}
\hat{O}_{6}^{0} = &   231\hat{l}_{z}^{6} -315l(l+1)\hat{l}_{z}^{4}+735\hat{l}_{z}^{4} %\nonumber \\
  %& 
  +105l^2(l+1)^2 \hat{l}_{z}^{2} -525l(l+1) \hat{l}_{z}^{2} + 294\hat{l}_{z}^{2}- \nonumber \\
  & 
  -5l^3(l+1)^3 + 40l^2(l+1)^2 - 60l(l+1),   \label{O60}\\
\hat{O}_{6}^{4} = &  \frac{1}{4}\left\{\left(11\hat{l}_{z}^{2} - l(l+1) -38\right)
  \left(\hat{l}_{+}^{4} +\hat{l}_{-}^{4} \right) %\right. \\
  + %& \left.
  \left(\hat{l}_{+}^{4} +\hat{l}_{-}^{4} \right)\left(11\hat{l}_{z}^{2} - l(l+1) -38\right)\right\} .
  \label{O64} 
\end{align}

\section{Choice of the energy window}\label{App:chewin}
\begin{figure*} [h!]%[htb]%%%%%%%%%%%%%%%%%%%%%%%%%%%%%%%%%%%%%%%%%%%%%%%%%%%%%%%%%%%%%%%
\includegraphics[%draft, 
width=0.32\textwidth]{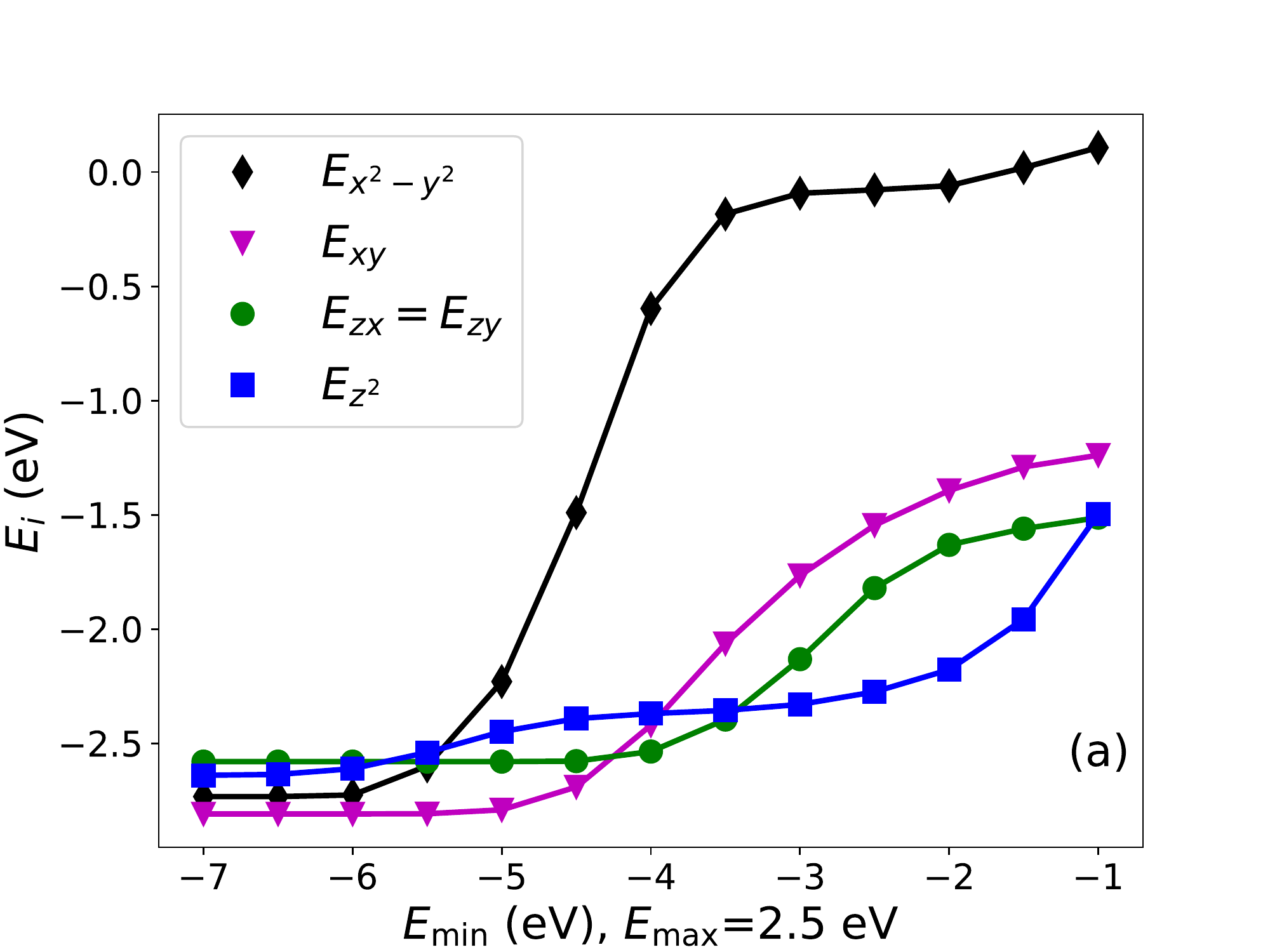}
\includegraphics[%draft, 
width=0.32\textwidth]{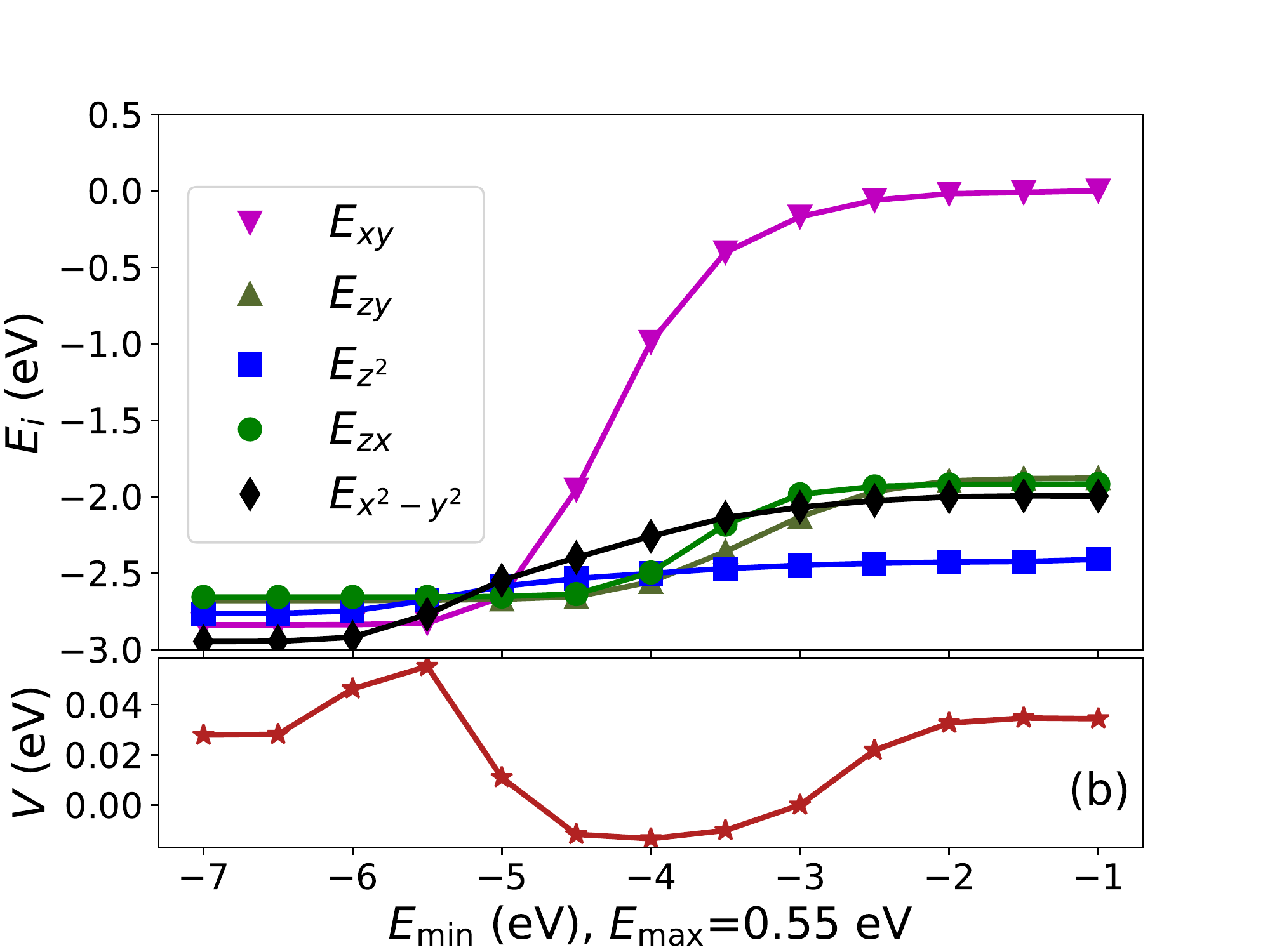}
\includegraphics[%draft, 
width=0.32\textwidth]{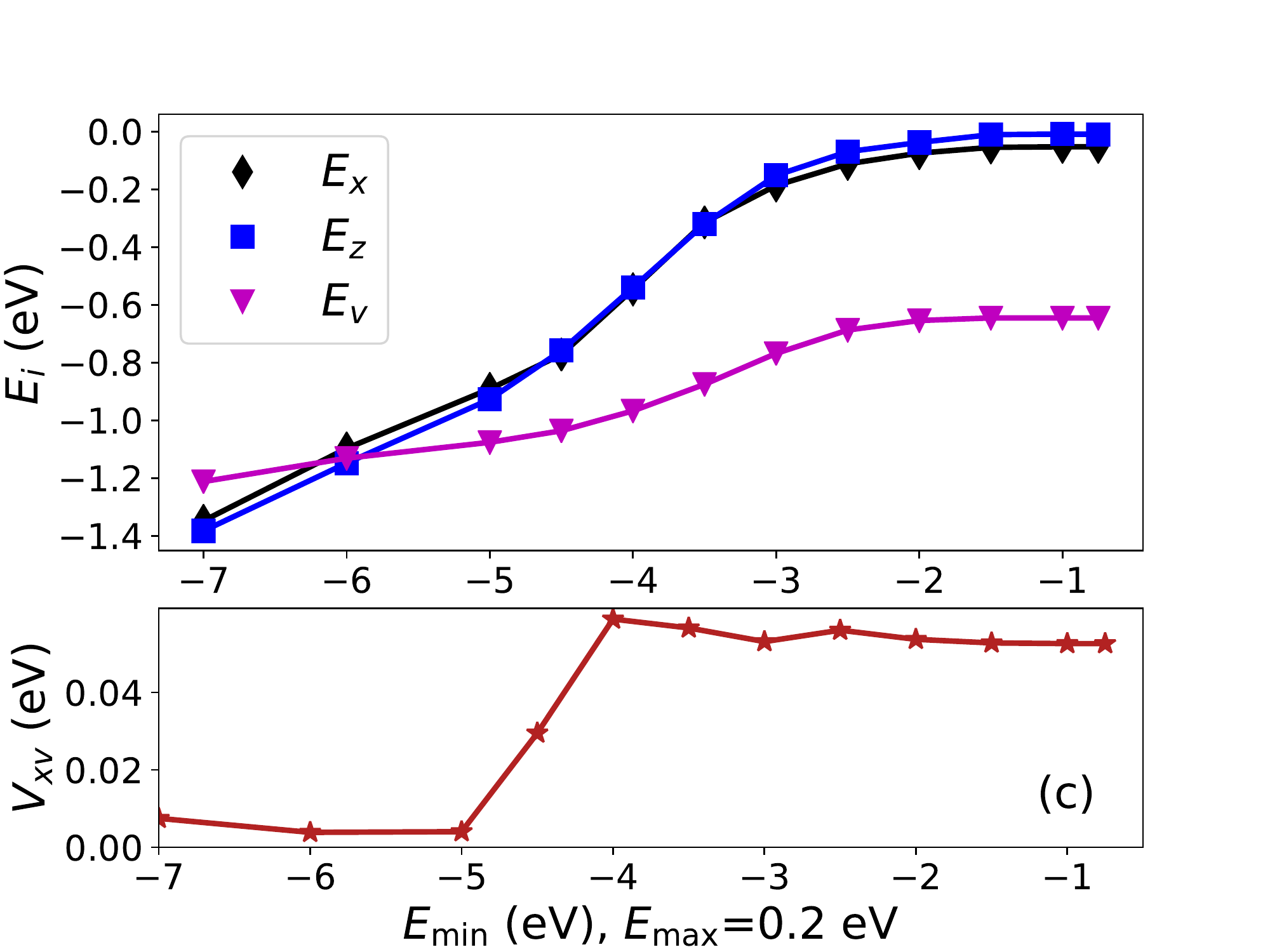}
\caption{\label{ewin}
The dependence of on-site energies of Wannier functions having $d$-function
symmetry on the energy window for the 
%high-$T_c$ parent 
compounds $\mathrm{CaCuO_2}$ (a, left), $\mathrm{Li_2CuO_2}$ (b, middle) 
and $\mathrm{ZnO:Co}$ (c, right). 
The energy window is defined 
by a function being 1 between $E_{min}$ and $E_{max}$ and falling of as a Gaussian with a width $dE=1$~eV outside this window for (a) and (b) and with a width $dE=0.5$~eV for (c). The lower limit of the energy window which was really taken into account in our determination of 
LF parameters was -3 eV for a and b, and -0.75 eV for c.}
\end{figure*} %%%%%%%%%%%%%%%%%%%%%%%%%%%%%%%%%%%%%%%%%%%%%%%%%%%%%%%%%%%%%%%%%%%%%
Figure \ref{ewin} illustrates the dependence of $d$-levels splitting 
on the width of the Wannier functions energy window. The upper limit of the 
window is set to the top of the valence band for each compound. 
When we chose the window that includes only antibonding states,
the splitting is much larger as compared with the case when the window includes 
all the states in the valence band. In the latter case the splitting is due 
to the electrostatic potential only (the bare crystal field), whereas in the former case 
it also includes the hybridization contribution into the ligand field. The variation of on-site energies is very small for ZnO:Co when we vary the lower bound of the 
energy window between -0.75 eV and - 2 eV since we are then in a gap of the density of states. That is not the case for the two cuprates where we choose the 
lower bound to be -3 eV separating in such a way antibonding and bonding states. However, there is no gap around -3 eV and we estimate an error of the 
on-site energies $E_i$ of about 0.1 eV. The Gaussian width $dE$ should be chosen as small as possible to reduce the error on $E_i$ but a too small value of 
$dE$ reduces the localisation of the Wannier orbitals. We have checked that the chosen values of $dE$ are the optimal compromise between precision of $E_i$ 
and localisation of Wannier orbitals.

\section{Dipole matrix elements of optical transition}
\label{Tanabe}
In this Appendix we use the parity breaking perturbation (\ref{pbt}) and 
%is then expressed as 
%\begin{equation}
%V_\st{odd}=B_3 \left(  \sqrt{\frac{5}{2}}Y_3^0 + Y_3^{+3} - Y_3^{-3}\right)
%\end{equation}
%\begin{eqnarray*}
%    H &=& V_3 + V_4\\
%    H &=& B_3 \left(  \sqrt{\frac{5}{2}}Y_3^0 + Y_3^{+3} - Y_3^{-3}\right) + B_4 \left( O_4^0 + 20 \sqrt{2} O_4^3 \right)
%\end{eqnarray*}\\
follow \cite{Sugano70} to express the dipole operator in the coordinate system
%\begin{eqnarray*}
% \nonumber to remove numbering (before each equation)
\begin{equation}
  \vec{k}_+= - \frac{1}{\sqrt{2}} \left( \vec{e}_x+ i \vec{e}_y \right) \; , \quad
  \vec{k}_- = + \frac{1}{\sqrt{2}} \left( \vec{e}_x- i \vec{e}_y \right) \; , \quad
  \vec{k}_0 = \vec{e}_z \; ,
%\end{eqnarray*}
\end{equation}
to be
\begin{equation}
\vec{P} = - P_+ \vec{k}_- - P_+ \vec{k}_+ + P_0 \vec{k}_0 \; ,
\end{equation}
%
%\begin{eqnarray*}
%    \overrightarrow{P} &=& q.r.\left(  -n_+ \overrightarrow{k_-} - n_- \overrightarrow{k_+} + 
%n_0 \overrightarrow{k_0}  \right)\\
%    \overrightarrow{P} &=& -e.r.\sqrt{\frac{4 \pi}{3}} \left(  - Y_1^{+1} .\overrightarrow{k_-} -  
%Y_1^{-1} .\overrightarrow{k_+} + Y_1^0. \overrightarrow{k_0}  \right)\\
%    \overrightarrow{P} &=&  - P_+ .\overrightarrow{k_-} -  P_- .\overrightarrow{k_+} + P_0. \overrightarrow{k_0}  \\
%\end{eqnarray*}
where the component $P_0 \propto Y_1^0$ and $P_{\pm} \propto Y_1^{\pm 1}$.
Using the rules for coupling two angular momenta we can calculate the dipole matrix element
for $\pi$-polarization
\begin{equation}
    W_0 \propto \bigg{|} \langle a |  \left(  \frac{9}{\sqrt{14}} Y_2^0 + \sqrt{ \frac{40}{7}} Y_4^0 + Y_4^{+3} - Y_4^{-3}    \right) |b \rangle \bigg{|}^2 \; ,\\
\end{equation}\\
 and for $\sigma$-polarization
\begin{equation}
   W_{\pm} \propto \bigg{|} \langle a |   \left( - \frac{3\sqrt{3}}{\sqrt{2}} Y_2^{\pm 1} + 5 Y_4^{\pm 1}  \pm \sqrt{28}   Y_4^{\pm 4}
    \pm 3 \sqrt{3} Y_2^{\mp 2} \mp  Y_4^{\mp 2}  \right) |b \rangle \bigg{|}^2 \; .\\
\end{equation}\\

\section{On-site Hamiltonian $H_{ii}$  for the actinide oxyde $\mathrm{UO_2}$}
\begin{table*}[h!]
\caption{\label{tab:HU} The matrix of the on-site Hamiltonian (in meV) using the basis of real spherical harmonics for the Wannier
expansion} of $\mathrm{UO_2}$. 
\begin{tabular}{r|ccccccc}
\hline 
 & $\left|5f,-3\right\rangle $ & $\left|5f,-2\right\rangle $ & $\left|5f,-1\right\rangle $ 
 & $\left|5f,0\right\rangle $  & $\left|5f,+1\right\rangle $ 
 & $\left|5f,+2\right\rangle $ & $\left|5f+3\right\rangle $ \\
\hline 
$\left\langle 5f,-3\right| $  & 365.58& 0& 36.62& 0&  & 0& 0\\
%\hline 
$\left\langle 5f,-2\right| $  & 0& 1212.03& 0& 0&  0& 0& 0\\
%\hline 
$\left\langle 5f,-1\right| $  & 36.62& 0& 346.67& 0& 0 & 0& 0\\
%\hline 
$\left\langle 5f,0\right| $   & 0& 0& 0& 393.94&  0& 0& 0\\
%\hline 
$\left\langle 5f,+1\right| $  & 0& 0& 0& 0&  346.67& 0& -36.62\\
%\hline 
$\left\langle 5f,+2\right| $  & 0& 0& 0& 0& 0 & 318.30& 0\\
%\hline 
$\left\langle 5f+3\right| $   & 0& 0& 0& 0&  -36.62& 0& 365.58
\end{tabular}
\end{table*}

\end{widetext}

\bibliography{w17}
\end{document}